\documentclass[aps,prd,longbibliography]{revtex4-1}
\makeatletter
\let\l@en\l@english
\makeatother
\usepackage[english]{babel}
\bibpunct{[}{]}{;}{n}{}{}
\usepackage{hyperref}
\usepackage{amsmath}
\usepackage{graphics}
\usepackage{graphicx}
\usepackage{epstopdf}

\usepackage{caption}
\usepackage{subcaption}
\usepackage{wrapfig}

\renewcommand{\baselinestretch}{1.5}

\newcommand{\ds}{\displaystyle}
\newcommand{\beq}{\begin{eqnarray}}
\newcommand{\eeq}{\end{eqnarray}}
\newcommand{\be}{\begin{equation}}
\newcommand{\ee}{\end{equation}}
\newcommand{\beqq}{\begin{eqnarray*}}
\newcommand{\eeqq}{\end{eqnarray*}}

\begin{document}


\title{Direct reconstruction of two-dimensional currents in thin films from magnetic field measurements}
\author{Alexander Y. Meltzer}
\thanks{These authors had equal contribution to this work;
Correspondence should be addressed to alexander.meltzer@weizmann.ac.il}
\affiliation{Department of Condensed Matter Physics, Weizmann Institute of Science, Rehovot 76100, Israel}
\author{Eitan Levin$^*$}
\affiliation{Department of Condensed Matter Physics, Weizmann Institute of Science, Rehovot 76100, Israel}
\author{Eli Zeldov}
\affiliation{Department of Condensed Matter Physics, Weizmann Institute of Science, Rehovot 76100, Israel}

\date{\today}

\begin{abstract}
Accurate determination of microscopic transport and magnetization currents is of central importance for the study of the electric properties of low dimensional materials and interfaces, of superconducting thin films and of electronic devices. Current distribution is usually derived from the measurement of the perpendicular component of the magnetic field above the surface of the sample, followed by numerical inversion of the Biot-Savart law. The inversion is commonly obtained by deriving the current stream function $g$, which is then differentiated in order to obtain the current distribution. However, this two-step procedure requires filtering at each step and, as a result, oversmoothes the solution. To avoid this oversmoothing we develop a direct procedure for inversion of the magnetic field that avoids use of the stream function. This approach provides enhanced accuracy of current reconstruction over a wide range of noise levels. We further introduce a reflection procedure that allows for the reconstruction of currents that cross the boundaries of the measurement window. The effectiveness of our approach is demonstrated by several numerical examples.

\end{abstract}

\maketitle
\section{Introduction}

Determination of two-dimensional current distribution from measurement of the normal component of a magnetic field is an important and commonly used tool for the investigation of a wide range of physical systems including high temperature superconductors \cite{Feldman2, Jooss1, JoossN, JRecPashitski17011997,HTM1,Sun2014}, topological states of matter \cite{SCANNING61,Spanton2014}, oxide heterostructures \cite{Band1,Band2}, carbon nanotubes and nanostructures \cite{Chang,Shadmi, SOT5}, as well as for the nondestructive evaluation of semiconductor circuits \cite{Fleet1999}. Mapping of the local magnetic fields is commonly attained by scanning Hall probes \cite{Feldman2,Bending1999,Kirtley2010,Grigorenko2001,hp1,hp2,Marchiori2017}, Hall-probe arrays \cite{HPA2,HPA3}, magneto-optical imaging (MOI) \cite{JoossN, JRecPashitski17011997, MO1,MO2,MO3,MO4,MO5,MO6,MO7,MO8,MO9,Jooss1}, and scanning superconducting quantum interference devices (SQUIDs) \cite{comp2,comp1,SCANNING2,SCANNING4, SCANNING6, Hykel2014, SCANNING9, Kirtley2010,Kremen2016,SCANNING61}. These techniques generate micrometer-to-millimeter scale two-dimensional images of the normal component of the magnetic field $B_z(x,y)$ above a sample. Recently however, nanoscale magnetic imaging has become a rapidly developing area of metrology, based on technological advances in scanning nitrogen vacancy (NV) centers in nano-diamonds \cite{Maletinsky, NV1,NV2,Chang, Dovzhenko}, nano-SQUIDs \cite{Granata2015, SOT1, SCANNING3, SOT2, SCANNING8, Lachman2015, SCANNING1, SCANNING5, Aviram}, and cold atom chips \cite{AC1, Aigner, Yang}. These techniques have the potential to provide higher spatial resolution, nanoscale proximity to the sample surface, improved field sensitivity, and lower measurement noise.

To take advantage of these new developments in nanoscale magnetic imaging, accurate analytical methods for the reconstruction of electric currents are required. Reconstruction of current distribution from the measured out-of-plane magnetic field requires inversion of the Biot-Savart law, which poses a number of challenges \cite{ Jooss1,exactFT,JoossN,CG-FFT,CG,Feldman,Feldman2}. First, the inversion equation, formulated as a Fredholm integral equation of the first kind, is ill-posed, resulting in amplification of the high spatial frequency components during the inversion process. In fact, high frequencies are never negligible in practice and therefore dominate the solution unless damped during the inversion \cite{MMLavrentiev1967}. Thus a na\"{\i}ve inversion of the Biot-Savart law is unstable and must be regularized. The second complication arises from the long-range nature of current-induced magnetic fields. The magnetic field in the imaged area can be affected by currents flowing outside the field of view, making the inversion equation underdetermined. Therefore, in order to obtain an accurate and unique solution one must make assumptions about the behavior of the current outside the measurement window. This problem is usually resolved by assuming that the entire current distribution is encompassed in the measurement window, similarly to the case where magnetization currents flow in closed loops. However, in the case of externally applied transport currents that significantly contribute to the measured field and necessarily cross the boundaries of the imaged area, this assumption is invalid and does not even constitute a good approximation.

Various approaches that address the instability of the inversion of the Biot-Savart law have been utilized so far. These introduce additional control parameters such as a cutoff frequency \cite{exactFT,JoossN} or a limitation on the number of numerical iterations \cite{CG-FFT,CG}. However, none of these methods provide systematic means for determining the optimal control parameters, with the exception of Feldmann \cite{Feldman} who recognizes the inversion problem as mathematically ill-posed. To overcome this difficulty Feldmann uses the Tikhonov regularization scheme \cite{tikhonov1977solutions} for current reconstruction, in which a free regularization parameter is used for controlling the smoothness of the solution. He then applies the Generalized-Cross Validation (GCV) method \cite{wahba1990spline} for methodical determination of the regularization parameter. However, the method of Feldmann, as presented in \cite{Feldman}, is not accurate at low heights above the sample. This is particularly disadvantageous for the next generation techniques, which aim to provide magnetic imaging for current reconstruction at nanometer heights above the sample surface in order to improve the sensitivity and the spatial resolution \cite{Chang, SOT1, Yang}.

The instability problem in the commonly used inversion methods for the Biot-Savart law is further exacerbated by the use of an auxiliary stream function for the inversion \cite{JoossN,CG,CG-FFT,Feldman}. Specifically, these methods determine the current distribution $\textbf{J}(x,y)$ in the sample by a two-step procedure. First, the stream function $g(x,y)$ is derived by inversion from the measured magnetic field $B_z(x,y)$ and the current is then determined from the $g$ function using the relation
\be \label{J}
\textbf{J}=\nabla\times[g(x,y)\widehat{z}].
\ee
In these two-step inversion methods (which we term GI methods), each of the steps is able to amplify the noise. In the first step, the noise is controlled by a regularization procedure that filters high spatial frequencies from the reconstruction. The resulting reconstructed $g$, however, is usually not sufficiently smooth to be differentiated with a regular numerical differentiation, which significantly amplifies any remaining noise. Consequently, it is necessary to apply an additional smoothing filter to $g$, such as the Savitzky-Golay filter proposed by Feldmann \cite{Feldman}, prior to differentiation. Application of a second filter in addition to the Tikhonov filter results in the smoothing of fine details in the solution that would have been otherwise preserved. Preservation of fine details is important in many cases. One such example is the reconstruction of sharp one-dimensional paths of higher current density at oxide interfaces \cite{Band1}, where over-smoothing can lead to inconclusive results. Exception from the above methodology, in which a one-step procedure is used, was proposed in \cite{exactFT}. However, the solution of the direct problem given in this paper differs from ours and does not use a rigorous regularization.

As mentioned above, the presence of external contributions to the magnetic field makes the inversion procedure more difficult. As far as we know, this problem was not addressed systematically before and all the methods cited above assume that the entire current distribution is contained in the measurement window. This is a severe restriction for most experimental setups, even in the absence of an external transport current that requires the use of an enlarged measurement window to ensure enclosure of all the currents. In the present work, the instability of the inverse problem and the presence of currents crossing the image boundary, which challenge the magnetic field inversion schemes, are addressed by the introduction of a number of novel procedures, as detailed below.

(i) We introduce a new inversion method utilizing the Tikhonov regularization in which the current distribution is obtained through a single step inversion of the measured $B_z(x,y)$, without the need for the intermediate derivation and differentiation of the stream function $g$. We show that this direct inversion (DI) scheme provides substantial improvement in the accuracy of current reconstruction over a wide range of noise levels. We also find that the quality of the reconstructions is not very sensitive to the exact value of the imaging height above the sample. This property is important as the exact height is usually not known in practice.

(ii) We develop two systematic procedures to determine the free Tikhonov regularization parameter in conjunction with the DI method based on GCV (DI-GCV) and on Stein's Unbiased Risk Estimate (SURE) \cite{Stein1981,Ramani2008,ProjectedSure} (DI-SURE). For $B_z(x,y)$ imaged at low heights $h$ above the sample the two procedures give comparable results, however at larger $h$ DI-SURE is preferable.

(iii) We introduce a reflection procedure addressing the transport current challenge. By symmetrically extending the $B_z(x,y)$ image we show that a reliable inversion can be attained in the case of currents crossing the boundary of the magnetic image. This reflection procedure performs best in conjunction with the DI-SURE regularization.

(iv) We improve the existing GI-GCV method and develop an alternative GI-SURE method, both of which can handle transport currents.

(v) The four schemes DI-GCV, DI-SURE, GI-GCV, and GI-SURE are applied to solve specific numerical examples, and their solutions are analyzed and compared showing the advantages and limitations of each method.

(vi) A user-friendly code is provided for all four inversion methods \footnote{A MATLAB-based implementation of the described algorithms and the numerical examples used in this paper can be found at:  {https://www.weizmann.ac.il/condmat/superc/software/}.}.

This paper is organized as follows. In Sec. \ref{Sec2} we briefly describe the GI method. In Sec. \ref{Sec3} we present our DI method for current reconstruction. In order to recover currents crossing the image boundary we introduce the reflection scheme in Sec. \ref{rbc} and the DI-SURE method in Sec. \ref{RegParSec}. In Sec. \ref{Numres} we present and discuss numerical results of two-dimensional current reconstruction using the GI and the DI methods. A new algorithm for noise variance estimation required for the SURE parameter-choice method is presented in the Appendix.

\section{Stream function GI method \label{Sec2}}
\subsection{Forward problem}
We begin by defining the problem and summarizing the stream function GI method \cite{Feldman}. The current $\textbf{J}$ flows in a three-dimensional thin film of thickness $d$, bounded in space by $-d/2\leq z\leq d/2$, $0\leq x \leq w$ and $0\leq y \leq l$. The measurement plane is parallel to the surface of the sample and to the $xy$-plane. Inside the film $\textbf{J}$ is static, depends on $x$ and $y$ and is uniform along the z-axis inside the sample. The magnetic field is measured at height $h=z-d/2$ above the sample, where $z$ is the $z$-coordinate of the measurement plane. We assume the field detector to be sensitive only to the $z$ component of the magnetic field and small enough, so that its nonzero sensing area does not distort the reconstructed current.

The experimentally measured field distribution $B_z(x,y)$ at height $z$ is related to the true currents in sample $J_{true}$ and their corresponding stream function $g_{true}$ through
\be\label{Eq1Conv}
B_z(x,y,z) = K(x,y,z)\ast g_{true}(x,y)+N(x,y),
\ee
where $N(x,y)$ is an additive noise of zero mean and constant variance $\sigma^2$, the kernel $K$ can take different forms depending on the assumptions of the problem, and the convolution of $F(x,y)$ and $f(x,y)$ is given by
\be
F(x,y)\ast f(x,y) = \int\limits_{-\infty}^{\infty}\int\limits_{-\infty}^{\infty}F(x-x',y-y')f(x',y')dx'dy',
\ee

For reconstruction of volume currents in a thin film with a non-negligible thickness $d$ the kernel is given by
\be\label{gKer}
K(x,y,z,d) = \frac{\mu_0}{4\pi}\left( \frac{z-d/2}{[x^2+y^2+(z-d/2)^2]^{3/2}}-\frac{z+d/2}{[x^2+y^2+(z+d/2)^2]^{3/2}}\right),
\ee
where $\mu_0$ is the permeability of free space. We define the two-dimensional Fourier transform and its inverse as
\beq
\label{FTandIFT1}
\widehat{f}(u,v) &=& \ds\int\limits_{-\infty}^{\infty}\int\limits_{-\infty}^{\infty}f(x,y)e^{-2\pi i(ux+vy)} dxdy, \\
f(x,y) &=& \ds\int\limits_{-\infty}^{\infty}\int\limits_{-\infty}^{\infty}\widehat{f}(u,v)e^{2\pi i(ux+vy)} dudv, 
\eeq
respectively, abbreviated as $\widehat{f}=\mathcal{F}[f]$ and $f=\mathcal{F}^{-1}[\widehat{f}]$.
The Fourier transform of Eq.\eqref{gKer} can be evaluated analytically as
\be\label{FTgKer}
\widehat{K}(u,v,z,d) = \mu_0 e^{-2\pi\sqrt{u^2+v^2}z}\sinh(\pi d\sqrt{u^2+v^2}).
\ee
If $d\ll h$, we can use the concept of sheet current that assumes that currents are confined to an infinitesimally thin film, with the corresponding kernel given by
\be\label{gKerSurfaceCurr}
K(x,y,z) = \frac{\mu_0}{4\pi}\frac{2z^2-x^2-y^2}{(x^2+y^2+z^2)^{5/2}},
\ee
and its Fourier transform given by
\be\label{FTgKerSurfaceCurr}
\widehat{K}(u,v,z) = \mu_0\pi\sqrt{u^2+v^2}e^{-2\pi\sqrt{u^2+v^2}z}.
\ee
Both kernels \eqref{gKer} and \eqref{gKerSurfaceCurr} can be thought of as low-pass filters with a cutoff frequency governed by the imaging height $z$. As such, they make the problem of approximating $g_{true}(x,y)$ in Eq. \eqref{Eq1Conv} ill posed and requiring regularization for a proper reconstruction \cite{Feldman}.

Note that kernels \eqref{gKer} and \eqref{gKerSurfaceCurr} and their matching stream functions have different dimensions. In the case of a film of thickness $d$ the current density $\textbf{J}$ is given in units of A/m$^2$, $g$ in units of A/m, and $K$ in units of T/A$\cdot$m. In the case of sheet currents $\textbf{J}$, $g$, and $K$ are correspondingly given in units of A/m, A, and T/A$\cdot$m$^2$.

\subsection{Inverse problem \label{TIP1}}
Approximation of $g_{true}(x,y)$ in \eqref{Eq1Conv} by Tikhonov regularization for a measured magnetic field $B_z$, consists of finding the $g_\lambda$ that solves the problem
\be\label{Eq1ConvFunct}
\min_{g_\lambda} (||K\ast g_\lambda - B_z||_2^2 + \lambda||L g_\lambda||_2^2),
\ee
for a given regularization parameter $\lambda$ and regularization operator $L$, where the 2-norm is defined as
\be
||f(x,y)||_2^2 = \int\limits_{-\infty}^{\infty}\int\limits_{-\infty}^{\infty} |f(x,y)|^2 dxdy.
\ee
The regularization parameter $\lambda$ in \eqref{Eq1ConvFunct}  sets the balance between a solution dominated by noise for small $\lambda$ and an over-smoothed solution for large $\lambda$. In order to penalize non-smooth solutions we define $L = \nabla^2$ following \cite{Feldman}.  It can be shown that the minimizer of Eq.\eqref{Eq1ConvFunct} is given by
\be\label{1convRegSoln}
g_{\lambda}(x,y) = \mathcal{F}^{-1}\left[ \frac{\overline{\widehat{K}}(u,v)\widehat{B}_z(u,v)}{|\widehat{K}(u,v)|^2+\lambda (2\pi)^4(u^2+v^2)^2}\right],
\ee
where a bar denotes complex conjugation. The current distribution can be found similarly to (\ref{J}) using
\be \label{J2}
\textbf{J}_{\lambda}=\nabla\times[g_{\lambda}(x,y)\widehat{z}].
\ee
We note that the stream function $g_\lambda$ is defined up to a gradient term, whereas the current $\textbf{J}_{\lambda}$ is unique \cite{CG}.

The regularization parameter $\lambda$ can be estimated using the GCV method \cite{wahba1990spline}, which seeks to approximately minimize the Predictive Mean-Square Error (PMSE), $||K\ast g_{true}-K\ast g_\lambda||_2^2$, where $g_{true}$ is the unknown true stream function. Since $g_{true}$ is not known, the GCV method minimizes a function slightly different from the PMSE and is given by
\be\label{GCV1}
G_1(\lambda) = \frac{||B_z - K\ast g_\lambda||_2^2}{\left(T_1(\lambda)\right)^2},
\ee
where $T_1(\lambda)$ is the residual effective degrees of freedom used in regression analysis \cite[p. 63]{wahba1990spline} and in our case it formally equals
\be
T_1(\lambda) = \int_{-\infty}^{\infty} \int_{-\infty}^{\infty}1- \frac{|\widehat{K}(u,v)|^2}{|\widehat{K}(u,v)|^2+\lambda (2\pi)^4(u^2+v^2)^2}dudv.
\ee
A more intuitive presentation of the GCV method and its connection to the PMSE can be found in Ref. \cite{PP2016}.

\subsection{Numerical implementation \label{ni1}}
In practice the magnetic field is sampled on a rectangular grid with $N$ points in the $x$ direction distanced $\Delta x$ units apart and $M$ points in the $y$ direction distanced $\Delta y$ units apart. Thus, the physical space grid consists of the points $(n\Delta x,m \Delta y)$ for $n=0,1,...,N-1$ and $m=0,1,...,M-1$ and the frequency space grid of the points $(\frac{k}{N\Delta x},\frac{l}{M\Delta y})$ for $k=0,1,...,N-1$ and $l=0,1,...,M-1$. We can approximate equation \eqref{1convRegSoln} on the discrete grid by using the Discrete Fourier Transform (DFT) and its inverse, defined as
\beq \label{FFTdef}
\widehat{f}_{kl} &=& \Delta x\Delta y\sum_{n=0}^{N-1} \sum_{m=0}^{M-1} f_{nm} e^{-i2\pi kn/N-i2\pi lm/M},\\
\label{BFTdef}
 f_{nm} &=& \frac1{\Delta x\Delta y}\frac{1}{NM} \sum_{k=0}^{N-1} \sum_{l=0}^{M-1} \widehat{f}_{kl} e^{i2\pi kn/N+i2\pi lm/M},
\eeq
respectively, abbreviated as
\beq
\widehat{f}_{kl} &=& \text{DFT}\left[f_{nm}\right]_{kl},\\
 f_{nm} &=& \text{DFT}^{-1}\left[\widehat{f}_{kl}\right]_{nm}
\eeq
Then, we can approximate \eqref{1convRegSoln} as
\be\label{gnm}
g_{nm} = \text{DFT}^{-1}\left[\frac{\overline{\widehat{K}}_{kl}\widehat{B}_{kl}}{|\widehat{K}_{ kl}|^2+\lambda |\widehat{\nabla}^2_{kl}|^2}\right]_{nm},
\ee
where $\widehat{K}_{kl}=\widehat{K}(\frac{k}{N\Delta x},\frac{l}{M\Delta y},z)$ is defined in either \eqref{FTgKer} or \eqref{FTgKerSurfaceCurr}, the Laplacian is approximated by the second-order central finite difference and $\widehat{B}_{kl} = \text{DFT}[B_{nm}]_{kl}$. Note also that because \eqref{gnm} employs DFT for the inversion, it implicitly assumes periodic boundary conditions at the boundaries of the measurement window. In the presence of currents crossing the boundaries, this assumption leads to highly inaccurate reconstructions as discussed in Sect. \ref{rbc}, making this inversion method inapplicable in such cases.

The stream function $g_{nm}$ reconstructed from a noisy measurement of $B_{nm}$ is not smooth. Therefore estimation of electric current using (\ref{J2}) by a simple numerical differentiation is not accurate and will amplify any noise left in $g_{nm}$. A more appropriate method for computation of the derivatives in this case is the Savitzky-Golay differentiation filter \cite{pressnumerical} which fits a polynomial of degree $p$ to each set of $2q+1$ successive data points by least squares. In this paper the current $\textbf{J}_{\lambda}$ is estimated by the differentiation of the fitted polynomial, using $p=2$ and $q=2$, as suggested in \cite{Feldman}.

For the GI-GCV method, the regularization parameter $\lambda$ in \eqref{gnm} is found using the GCV scheme \eqref{GCV1}. The discrete version of the function $G_1(\lambda)$ is given by \footnote{Note, that the formula for GCV given in Equation 15 in \cite{Feldman}  contains a typographical error.}
\be\label{GCVF1}
G_1(\lambda) = \frac{\sum_{k=0}^{N-1}\sum_{l=0}^{M-1} (1-\widehat Z_{kl})^2|\widehat{B}_{kl}|^2}{\left({NM}-\sum_{k=0}^{N-1}\sum_{l=0}^{M-1} \widehat Z_{kl}\right)^2}.
\ee
where
\be\label{GCVF2}
\widehat Z_{kl} = \frac{|\widehat{K}_{kl}|^2}{|\widehat{K}_{kl}|^2 + \lambda|\widehat{\nabla}^2_{kl}|^2}.
\ee
The regularization parameter $\lambda$ is then estimated as the minimizer of the function $G_1(\lambda)$.

\begin{figure}
    \centering
    \includegraphics[width=.45\textwidth]{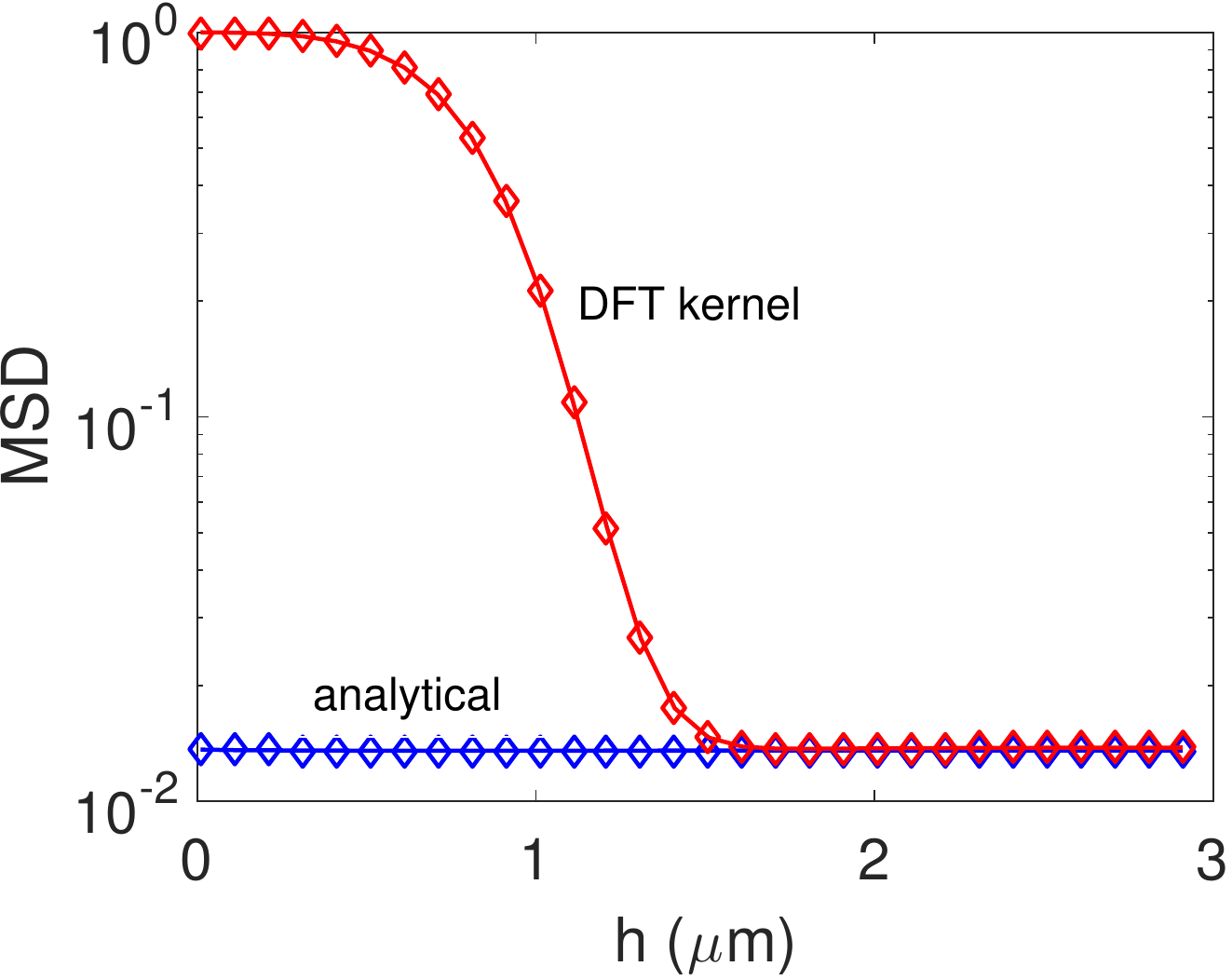}
    \caption{The MSD of current reconstruction vs. the field measurement height $h$, obtained using the analytical kernel \eqref{FTgKer} and the DFT kernel \eqref{DFTK}. The calculation was performed on sample A (as described in Sect. \ref{Numres}) with no additive noise, using the GI-GCV scheme with grid size $\Delta x=1~\mu \text{m}$. The DFT kernel becomes accurate only for $h > 1.7~\mu \text{m}$.}
\label{fig:KerMSD}
\end{figure}

It is important to note that evaluation of the kernel using the discrete transform
\be \label{DFTK}
\widehat{K}_{kl}=\text{DFT}[K(n\Delta x,m\Delta y,z)]_{kl},
\ee
as suggested in \cite{Feldman}, should be avoided due to the large inaccuracy of this approximation, compared to the exact expressions in \eqref{FTgKer} and \eqref{FTgKerSurfaceCurr}, especially for small heights. This is demonstrated in Fig. \ref{fig:KerMSD}, where we measure the accuracy of current reconstruction in absence of noise by the mean square deviation (MSD) defined as
\be\label{MSD}
\begin{aligned}
\text{MSD} &=& &\frac{||\textbf{J}_{true}(x,y)-\textbf{J}_{\lambda}(x,y)||_2^2}{||\textbf{J}_{true}(x,y)||_2^2} \\ &=& &\frac{||j_{x,true}-j_{x,\lambda}||_2^2+||j_{y,true}-j_{y,\lambda}||_2^2}{||j_{x,true}||_2^2+||j_{y,true}||_2^2}.
\end{aligned}
\ee
Here $\textbf{J}_{true}$ is the actual current in sample A (as described in Sect. \ref{Numres}) and $\textbf{J}_\lambda$ is the current reconstructed from the calculated magnetic field at height $h$ using either the analytical kernel \eqref{FTgKer} or the DFT kernel \eqref{DFTK}. Figure \ref{fig:KerMSD} shows that the DFT kernel introduces a large error and cannot be used for heights smaller than about twice the grid spacing, which in this example was $\Delta x=1$ $\mu$m.

To summarize, the GI method constitutes inverting the magnetic field in the following steps:
\begin{enumerate}
  \item Estimate the regularization parameter $\lambda$ as the minimizer of $G_1(\lambda)$ in \eqref{GCVF1}.
  \item Compute the stream function $g(\lambda)$ using \eqref{gnm}.
  \item Obtain the currents $\textbf{J}_{\lambda}$ by applying the Savitsky-Golay differentiation filter to $g$ as described above with $p=q=2$.
\end{enumerate}
In the following Section, we develop a new method that does not require the intermediate computation of the stream function $g$.

\section{The direct inversion (DI) of the Biot-Savart law \label{Sec3}}
In this section we introduce an alternative formulation of the inversion problem, which produces higher quality reconstructions, in particular in the presence of low noise. In addition, we present both a reflection procedure for the reconstruction of currents crossing the image boundaries as well as a projected SURE method for determination of the regularization parameter in this case.
\subsection{The forward problem}
The forward problem of calculating the magnetic field, given distribution of the current, requires the solution of the Biot-Savart law, which is given by
\be \label{BSdef}
\textbf{B}(\textbf{r}) = \frac{\mu_0}{4\pi}\int_{\Omega} \frac{\textbf{J}(\textbf{r}')\times(\textbf{r}-\textbf{r}')}{|\textbf{r}-\textbf{r}'|^3}d\textbf{r}' ,
\ee
where $\textbf{r}$ is the observation coordinate, $\textbf{r}'$ is the source coordinate and $\textbf{J} = j_x(x,y)\widehat{x}+j_y(x,y)\widehat{y}$ is the current density vector field. The $z$-component of the magnetic field can be found from \eqref{BSdef} as
\be
\label{eq:BiotSavartZcomp} B_z(x,y,z) = \frac{\mu_0}{4\pi}\int\limits_{-\frac{d}{2}}^{\frac{d}{2}}\int\limits_{0}^{w}\int\limits_{0}^{l} \frac{(y-y')j_x(x',y')-(x-x')j_y(x',y')}{[(x-x')^2+(y-y')^2+(z-z')^2]^{3/2}} dx'dy'dz' .
\ee
We can rewrite Eq.\eqref{eq:BiotSavartZcomp} as
\beq B_z(x,y,z) &=& \int\limits_{-d/2}^{d/2} [A_1(x,y,z-z')\ast j_x(x,y) + A_2(x,y,z-z')\ast j_y(x,y)]dz' \nonumber\\
&=& K_1(x,y,z,d)\ast j_x(x,y) + K_2(x,y,z,d)\ast j_y(x,y), \label{conv2Eq}
\eeq
where kernels $A_1$ and $A_2$ are given by
\beq
A_1(x,y,z) &=& \frac{\mu_0}{4\pi}\frac{y}{[x^2+y^2+z^2]^{3/2}} , \label{eq:a1a}\\
A_2(x,y,z) &=& \frac{\mu_0}{4\pi}\frac{-x}{[x^2+y^2+z^2]^{3/2}} ,\label{eq:a1b}
\eeq
and kernels $K_1$ and $K_2$, which depend on the thin film thickness $d$, are given by
\beq
K_1(x,y,z,d) &=& \left\{\begin{array}{ll}
\frac{\mu_0}{4\pi}\frac{y}{x^2+y^2}\left[\frac{z+d/2}{\sqrt{x^2+y^2+(z+d/2)^2}}- \frac{z-d/2}{\sqrt{x^2+y^2+(z-d/2)^2}}\right],& x^2+y^2>0,\\ 0 & x^2+y^2=0,
\end{array}\right.
\label{zIntK12a}\\
K_2(x,y,z,d) &=& \left\{\begin{array}{ll}
\frac{\mu_0}{4\pi}\frac{x}{x^2+y^2}\left[\frac{z-d/2}{\sqrt{x^2+y^2+(z-d/2)^2}}- \frac{z+d/2}{\sqrt{x^2+y^2+(z+d/2)^2}}\right],& x^2+y^2>0,\\ 0 & x^2+y^2=0,
\end{array}\right. \label{zIntK12b}
\eeq
for $|z|>d/2$. The analytical Fourier transforms of kernels $A_1$ and $A_2$ are
\beq
\label{FTa1a}
\widehat{A}_1(u,v,z) &=& \left\{\begin{array}{ll} -i\frac{\mu_0}{2} e^{-2\pi\sqrt{u^2+v^2}z}\frac{v}{\sqrt{u^2+v^2}},& u^2+v^2>0,\\ 0,& u^2+v^2=0, \end{array}\right.\\
\label{FTa1b}
\widehat{A}_2(u,v,z) &=& \left\{\begin{array}{ll}  i\frac{\mu_0}{2} e^{-2\pi\sqrt{u^2+v^2}z}\frac{u}{\sqrt{u^2+v^2}},& u^2+v^2>0,\\ 0,& u^2+v^2=0, \end{array}\right.
\eeq
while those of kernels $K_1$ and $K_2$ are
\beq
\label{FTK1K2a}
\widehat{K}_1(u,v,z,d)\ &=& \left\{\begin{array}{ll} -i\frac{\mu_0}{2\pi}e^{-2\pi\sqrt{u^2+v^2}z}\sinh(\pi d\sqrt{u^2+v^2})\frac{v}{u^2+v^2}, & u^2+v^2>0,\\ 0,& u^2+v^2=0, \end{array}\right.\\
\label{FTK1K2b}
\widehat{K}_2(u,v,z,d)\ &=& \left\{\begin{array}{ll}  i\frac{\mu_0}{2\pi}e^{-2\pi\sqrt{u^2+v^2}z}\sinh(\pi d\sqrt{u^2+v^2})\frac{u}{u^2+v^2}, & u^2+v^2>0,\\ 0,& u^2+v^2=0. \end{array}\right.
\eeq
If $h\gg d$ we can use the concept of sheet currents, in which case the magnetic field is given simply by
\be
B_z(x,y,z) = A_1(x,y,z)\ast j_x(x,y) + A_2(x,y,z)\ast j_y(x,y), \label{OurSurfaceCurr}
\ee
without an integral in the $z$ direction.

The relation \eqref{conv2Eq} (or \eqref{OurSurfaceCurr}) leads us to the following compatibility condition. By applying the Fourier transform to \eqref{OurSurfaceCurr} we can present the relation for the zero mode ($u=v=0$) as
\be \label{BAJ}
\widehat{B}_z(0,0,z) = \widehat{K}_1(0,0,z) \widehat{j}_x(0,0) + \widehat{K}_2(0,0,z) \widehat{j}_y(0,0).
\ee
Since $\widehat{K}_1(0,0,z)$ and $\widehat{K}_2(0,0,z)$ are zero, the value of $\widehat{B}_z(0,0,z)$ should also be zero. In the real space the condition requires the mean value of $B_z$ to be zero. Thus, the compatibility condition implies that the mean value of the currents cannot be deduced from the measured field since it does not contribute to this field. The reconstruction of the current is therefore possible only up to an additive constant that represents a uniform current in the physical space. On the positive side, \eqref{BAJ} implies that our reconstruction is insensitive to offsets in the magnetic field that are usually inflicted by external sources.

For a better understanding of the problem we can find the length scale which characterizes kernels \eqref{FTa1a} - \eqref{FTa1b} on a grid. In a simple case of \(\Delta x=\Delta y =\Delta\) the kernels become dependent on one parameter,  the ratio between the height \(z\) and pixel size \(\Delta\). For kernels \eqref{FTK1K2a} - \eqref{FTK1K2b} the same argument applies if the ratio between the height and the sample thickness \(d\) is kept constant. To see this, we can rewrite our kernels in Fourier space outside the origin as
\be
\widehat{A}_1(u,v,z) = -i\frac{\mu_0}{2} e^{-2\pi\sqrt{k^2+l^2}\frac{z}{N\Delta}}\frac{l}{\sqrt{k^2+l^2}}, \label{r1}
\ee
where $k=0,\ldots,N$, $l=0,\ldots, M$ and $\Delta$ is the grid spacing. From \eqref{r1} it is easy to see that for fixed $N$ and $M$  it is only $z/\Delta$ that determines the decay of the kernel and the corresponding spatial resolution of the reconstructed currents. This finding is important because the kernel decay determines the smoothing effect of the kernel and consequently the ill-conditioning and hence the difficulty of the reconstruction, as described in the next subsection.

\subsection{The inverse problem}
Equations \eqref{conv2Eq} and \eqref{OurSurfaceCurr} enable us to find the magnetic field from either the volume or the sheet current distribution within the sample. The corresponding derivation of the currents from $B_z$ thus requires solving the inversion problem with two kernels. This task may seem to be challenging and less controllable as compared to the hitherto used GI method that involves only a single kernel. However, it is in fact more accurate, as it does not require the second Savitsky-Golay filter used in the GI method, enabling thus  reconstruction of finer details. In the following we develop this new DI method and demonstrate its advantages.

Assuming an additive noise model as in \eqref{Eq1Conv} we can rewrite \eqref{conv2Eq} and \eqref{OurSurfaceCurr} as
\be
B_z(x,y) = K_1\ast j_x(x,y) + K_2\ast j_y(x,y)+N(x,y), \label{jAddN}
\ee
where $K_1$ and $K_2$ in \eqref{zIntK12a}-\eqref{zIntK12b} can be replaced with kernels $A_1$ and $A_2$ in \eqref{eq:a1a}-\eqref{eq:a1b} respectively if $h\gg d$. The inversion of the Biot-Savart law \eqref{jAddN} and determination of the currents $j_x(x,y)$ and $j_y(x,y)$ from \eqref{OurSurfaceCurr} given $B_z$ is ill-posed. Therefore, similarly to Sec. \ref{Sec2}, we solve this problem using Tikhonov regularization by minimization of the following functional
\be\label{Eq2ConvFunct}
\min_{j_x,j_y} (||K_1\ast j_x + K_2\ast j_y - B_z||_2^2 + \lambda\left(||L j_x||_2^2 + ||L j_y||_2^2\right)), \ee
where the same parameter $\lambda$ multiplies both penalty terms due to the lack of a directional preference in the problem, and we set $L=\nabla^2$ as in Sect. \ref{TIP1}. For simplicity, we suppressed in \eqref{Eq2ConvFunct} the dependence of the kernels and of the currents on the variables $x$ and $y$. The regularized solutions that minimize Eq.\eqref{Eq2ConvFunct} are given by
\beq
\label{2convRegSoln_a}
    j_{x}(\lambda) = \mathcal{F}^{-1}\left[ \frac{\overline{\widehat{K}}_1\widehat{B}_z}{|\widehat{K}_1|^2+|\widehat{K}_2|^2+\lambda (2\pi)^4(u^2+v^2)^2}\right], & \\
\label{2convRegSoln_b}
    j_{y}(\lambda) = \mathcal{F}^{-1}\left[ \frac{\overline{\widehat{K}}_2\widehat{B}_z}{|\widehat{K}_1|^2+|\widehat{K}_2|^2+\lambda (2\pi)^4(u^2+v^2)^2}\right]. &
\eeq
It is easy to check that the reconstructed current \eqref{2convRegSoln_a}-\eqref{2convRegSoln_b} satisfies
\be
\nabla \cdot\textbf{J}=0. \label{Jcont}
\ee
Due to the compatibility condition \eqref{BAJ} the dc components of the currents are not defined by \eqref{2convRegSoln_a}-\eqref{2convRegSoln_b}, and as discussed above, cannot be reconstructed from the measured field. Therefore, we set them to zero, which is equivalent to assuming no uniform current flowing in the measurement window.

Discretizing Eqs. \eqref{2convRegSoln_a}-\eqref{2convRegSoln_b}, as in the previous section, we obtain
\beq\label{D2conv_a}
j_{x, nm}(\lambda) = \text{DFT}^{-1}\left[\frac{\overline{\widehat{K}}_{1, kl}\widehat{B}_{kl}}{|\widehat{K}_{1, kl}|^2+|\widehat{K}_{2, kl}|^2+\lambda |\widehat{\nabla}^2_{kl}|^2}\right]_{nm},& \\
\label{D2conv_b}
j_{y, nm}(\lambda) = \text{DFT}^{-1}\left[\frac{\overline{\widehat{K}}_{2, kl}\widehat{B}_{kl}}{|\widehat{K}_{1, kl}|^2+|\widehat{K}_{2, kl}|^2+\lambda |\widehat{\nabla}^2_{kl}|^2}\right]_{nm},&
\eeq
where $\widehat{K}_{j, kl}=\widehat{K}_j(\frac{k}{N\Delta x},\frac{l}{M\Delta y},z)$ is given by the analytic expressions in either \eqref{FTa1a}-\eqref{FTa1b} or \eqref{FTK1K2a}-\eqref{FTK1K2b}, and the Laplacian is approximated by the central second-order finite difference stencil.

In the presence of currents crossing the boundaries, a na\"{i}ve application of Eqs. \eqref{D2conv_a}-\eqref{D2conv_b} fails to produce an accurate solution due to the the artifacts caused by the DFT, which assumes periodicity of the measured field, and due to the fact that these equations satisfy \eqref{Jcont} everywhere, including the boundary. To overcome this problem we apply a reflection rule to the measured magnetic field, as explained in the next subsection.

\subsection{Reflection rule \label{rbc}}

In this section we consider the inversion problem in the presence of currents flowing through the image boundary. An accurate reconstruction of the currents through the boundary requires knowledge of the magnetic field outside the imaged region. In absence of such knowledge, the inversion equation \eqref{jAddN} becomes underdetermined and does not have a unique solution. A na\"{\i}ve application of the DFT as in Eqs. \eqref{gnm} or \eqref{2convRegSoln_a}-\eqref{2convRegSoln_b} assumes periodic boundary conditions, extending the currents periodically to infinity. Since the measured field produced by currents that cross the boundary of the measurement window is in general non-periodic, application of periodic boundary conditions in this case creates a discontinuity at the boundary. This in turn causes Gibbs oscillations of the reconstructed current at the same boundary. In addition, the current conservation property \eqref{Jcont}, fulfilled by either \eqref{J} in the GI method or by  \eqref{2convRegSoln_a}-\eqref{2convRegSoln_b} in the DI method, forces an incorrect closure of the current loops inside the reconstruction window, if periodic boundary conditions are used. Thus, to handle current distributions extending beyond the measurement window, we must either supply information about the field outside the measurement window, which is not generally available, or implement more appropriate boundary conditions for the currents, which we develop in this section.

To implement more appropriate boundary conditions for the currents, we suggest to replace the image of the magnetic field measurement with an extended image, such that the magnetic field outside the measurement area is a mirror image of the field inside the boundaries. Specifically, we symmetrically extend $B_z$ by embedding it into a larger matrix $\widetilde B_z$, such that
\be\label{eq:flipsForData}
\widetilde B_z = \left(\begin{array}{cc} B_{rc} & B_c \\ B_r & B_z \end{array}\right),
\ee
where $B_c$ is obtained from $B_z$ by flipping its columns, $B_r$ is obtained by flipping the rows and $B_{rc}$ is obtained by flipping both. The solution is then obtained by substituting $\widetilde B_z$ in Eqs. \eqref{2convRegSoln_a} and \eqref{2convRegSoln_b} for $B_z$ and taking only the part of the Tikhonov solution in the original window.

The reflections in \eqref{eq:flipsForData} ensure a continuous flow of the current across the different boundaries of the image by closing currents outside the measurement window, as shown in the following analysis. For simplicity of presentation the analysis is carried out in the continuous space. We examine first the effect of the reflection upon the reconstruction using the GI method. Since the reconstructed currents given by \eqref{2convRegSoln_a}-\eqref{2convRegSoln_b} are translationally invariant due to their implicit periodic extension by the DFT, we consider here only two boundaries  $x=0$ and $y=0$, where the reflections in \eqref{eq:flipsForData} ensure $B_z(-x,y)= B_z(x,y)$ and $B_z(x,-y)= B_z(x,y)$ respectively. We also need to recall that the convolution of two odd or two even functions is even and the convolution of an odd and an even function is odd.

\begin{figure}
\centering
\includegraphics[width=.45\textwidth]{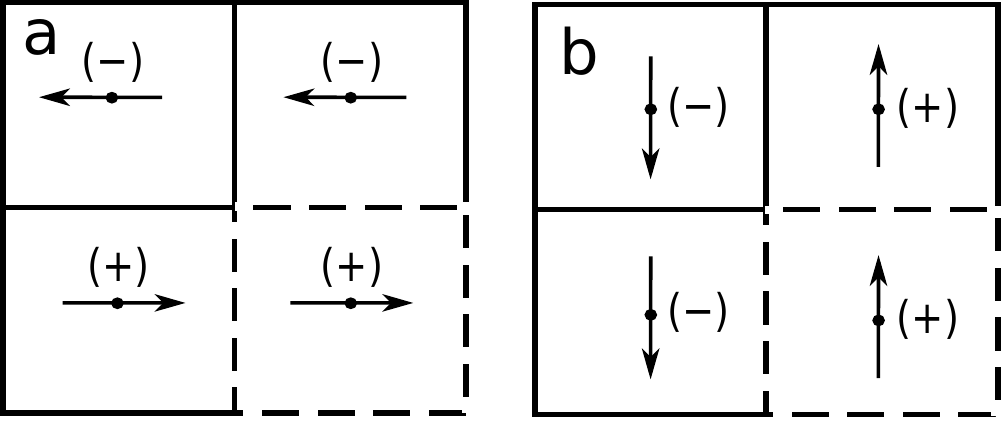}
\caption{A schematic description of the direction of currents (a) $j_x$ and (b) $j_y$ upon implementation of the symmetric extension of the field. The measurement window is marked by a dashed line.}
\label{fig:BCcurrElem}
\end{figure}

The kernel $K$, either from \eqref{gKer} or \eqref{gKerSurfaceCurr}, is even in both $x$ and $y$ directions. Disregarding the noise term in \eqref{Eq1Conv} we deduce that, if $B_z(-x,y)= B_z(x,y)$ then the function $g_{true}$ should also be even ($g_{true}(-x,y)= g_{true}(x,y)$) and, since the derivative of an even function is odd and vice versa, we obtain
\begin{align}
&&j_x(-x,y)=j_x(x,y),\quad\quad\quad\quad\quad\quad\quad\quad\quad\quad\label{jx1}\\
&&j_y(-x,y)=-j_y(x,y).\quad\quad\quad\quad\quad\quad\quad\quad\quad\quad\label{jy1}
\end{align}
On the other hand, if $B_z(x,-y)= B_z(x,y)$ then $g_{true}(x,-y)= g_{true}(x,y)$ and using the same argument we obtain
\begin{align}\label{jx2}
&&j_x(x,-y)=-j_x(x,y)\quad\quad\quad\quad\quad\quad\quad\quad\quad\quad\\
&&j_y(x,-y)=j_y(x,y). \quad\quad\quad\quad\quad\quad\quad\quad\quad\quad\label{jy2}
\end{align}
Therefore, upon crossing the boundary, the component of the current perpendicular to the boundary remains unchanged whereas the component parallel to the boundary changes its sign, as shown in Fig. \eqref{fig:BCcurrElem}.

To obtain similar results when the noise term is non-negligible we recall that the inverse Fourier transform of a real and even function is even and that of an imaginary and odd function is odd. Next, we rewrite \eqref{1convRegSoln} used for reconstruction of $g_{\lambda}$ as
\be\label{1convRegSoln2}
g_{\lambda}(x,y) = \mathcal{F}^{-1}\left[ \frac{\overline{\widehat{K}}(u,v)}{|\widehat{K}(u,v)|^2+\lambda (2\pi)^4(u^2+v^2)^2}\right]*{B}_z(x,y),
\ee
and since $\widehat{K}(u,v)$ is even and real, we conclude that $g_{\lambda}(-x,y)= g_{\lambda}(x,y)$ and $g_{\lambda}(x,-y)= g_{\lambda}(x,y)$ for $B_z$ even about $x=0$ and $y=0$ respectively. As a result \eqref{jx1}-\eqref{jy2} are satisfied by the Tikhonov solution $g_\lambda$.

Analysis of the effect of the symmetric extension on current reconstruction by the DI method is similar. Using either \eqref{conv2Eq} or \eqref{OurSurfaceCurr} and noting that $K_1$ and $A_1$ are even in $x$ and odd in $y$, while $K_2$ and $A_2$ are odd in $x$ and even in $y$ we deduce that if $B_z(-x,y)= B_z(x,y)$ then, using the aforementioned properties of convolution we obtain \eqref{jx1} and \eqref{jy1}. Similarly, if $B_z(x,-y)= B_z(x,y)$, we find that \eqref{jx2} and \eqref{jy2} are satisfied. This result is identical to the case of the GI method and is also illustrated by Fig. \ref{fig:BCcurrElem}. Taking the noise into account we use \eqref{2convRegSoln_a}-\eqref{2convRegSoln_b} for reconstruction of the current field. Applying a similar reasoning as above, we conclude that the currents obtained from \eqref{2convRegSoln_a}-\eqref{2convRegSoln_b} also satisfy \eqref{jx1}-\eqref{jy2}.

\subsection{Regularization parameter choice methods \label{RegParSec}}
In the present section we discuss the parameter choice methods for the reconstruction of current densities using Eqs. \eqref{D2conv_a}-\eqref{D2conv_b}. If the currents do not cross the boundaries we can still use the GCV method similar to the one discussed in Sec. \ref{TIP1}. The GCV for the DI method consists of minimization of the function
\be\label{G222}
G_2(\lambda) = \frac{||B_z - K_1\ast j_x(\lambda)-K_2\ast j_y(\lambda) ||_2^2}{\left(T_2(\lambda)\right)^2},
\ee
where $T_2(\lambda)$ is formally given by
\be
T_2(\lambda) = \int_{-\infty}^{\infty} \int_{-\infty}^{\infty}\left[1- \widehat{H}(\lambda)\right]\,du\,dv,
\ee
and
\be
\widehat{H}(\lambda) = \frac{|\widehat{K}_1(u,v)|^2+|\widehat{K}_2(u,v)|^2}{|\widehat{K}_1(u,v)|^2+|\widehat{K}_2(u,v)|^2+\lambda (2\pi)^4(u^2+v^2)^2}.
\ee
Similarly to \eqref{GCV1}, the function \eqref{G222} is designed so that its minimum is close to the minimum of the PMSE, which is defined by
$$\text{PMSE}(\lambda)=\|B_{true}-B(\lambda)\|_2^2,$$
where $B_{true}=B_z-N$ is the true value of the magnetic field, $
B(\lambda)=K_1*j_x(\lambda)+K_2*j_y(\lambda)$,
and $N$ is the unknown noise. The discrete version of the GCV method is given by
\be
G_2(\lambda) = \frac{\sum_{k=0}^{N-1}\sum_{l=0}^{M-1} (1-\widehat{H}_{kl}(\lambda))^2|\hat{B}_{kl}|^2}{\left(MN-\sum_{k=0}^{N-1}\sum_{l=0}^{M-1} \widehat{H}_{ kl}(\lambda)\right)^2},
\ee
where
\be
\widehat{H}_{ kl}(\lambda) = \frac{|\widehat{K}_{2, kl}|^2 + |\widehat{K}_{1, kl}|^2}{|\widehat{K}_{1, kl}|^2+|\widehat{K}_{2, kl}|^2+\lambda|\widehat{\nabla}^2_{kl}|^2}.
\ee

The values of $\lambda$ found using \eqref{G222} are typically very close to optimal. However, they become unsatisfactory when currents cross the boundaries of the measurement window. Even though utilization of the reflection rule, achieved by substituting $\tilde B_z$ (given by \eqref{eq:flipsForData}) for $B_z$ in \eqref{G222} provides a significant improvement, the regularization may still be far from optimal. More accurate estimates of $\lambda$ in this case can be obtained, however, by using the projected SURE method, which is similar to the method previously proposed in \cite{ProjectedSure,Ramani2008}. Particularly, let $P$ denote a projection operator from the enlarged domain $\widetilde{B}_z$ to a region inside $B_z$. For example, in our numerical tests we choose the image of the projection $P$ to contain the central 80\% of the measured field $B_z$. To find the regularization parameter which gives the best reconstruction we approximately minimize the projected PMSE norm  $f(\lambda)=\|P(\widetilde B_z-N-\widetilde B(\lambda))\|^2_2$, where $\widetilde B(\lambda)$ is calculated using the currents $(j_x(\lambda),j_y(\lambda))$ obtained by the inversion formulas \eqref{2convRegSoln_a}-\eqref{2convRegSoln_b} applied to the symmetrically extended magnetic field $\widetilde B_z$. We can rewrite $f(\lambda)$ as
\be\label{fl}
f(\lambda) = \|PN\|^2_2+\|P(\widetilde B_z- \widetilde B(\lambda))\|^2_2-2C(\lambda),
\ee
where, defining the $\ell_2$ inner product by $\langle\cdot{,}\cdot\rangle_2$, the last term is given by
\be\label{Rl}
C(\lambda)=\Re\left[ \left\langle PN,P(\widetilde B_z- \widetilde B(\lambda))\right\rangle_2\right].
\ee
The first term on the right hand side of (\ref{fl}) is independent of $\lambda$ and therefore can be neglected. The second term in (\ref{fl}) can be easily calculated, while $C(\lambda)$ cannot be exactly calculated due to its dependance on an unknown noise $N$. However it is possible to approximate $C(\lambda)$ as follows. First, we rewrite (\ref{Rl}) as
\be
C(\lambda) = \Re\left[ \left\langle PN,P(\widetilde B_{true}+N)\right\rangle_2\right] - \Re\left[\left\langle PN,P(B_{true}(\lambda)+ N(\lambda))\right\rangle_2\right],
\ee
where  $N(\lambda) = \mathcal{F}^{-1}\left[\widehat{H}(\lambda)\widehat{ N} \right]$, $B_{true}(\lambda) = \mathcal{F}^{-1}\left[\widehat{H}(\lambda)\widehat{\widetilde B}_{true}\right]$ and $\widetilde B_{true}$ is the symmetrically extended version of $B_{true}$.  We can then drop the terms $\Re[\langle PN,P\widetilde B_{true}\rangle_2]$ and $\Re[\langle PN,PB_{true}(\lambda)\rangle_2]$ as in \cite{PP2016,NearOpt2006} since their expected value is zero, so that
\be\label{Rnm}
C(\lambda) \simeq \widehat C(\lambda)\equiv\|PN\|^2_2 - \Re[\langle PN,PN(\lambda)\rangle_2].
\ee

In the discrete version of the projected SURE method we can approximate \eqref{Rnm} by replacing the unknown noise $N_{nm}$ with a known noise $N_{1,nm}$, having the same mean and variance as $N_{nm}$ \cite{Ramani2008}. Following \cite{StochEst} we choose $N_{1,nm}$ such that its components are either $+\sigma$ or $-\sigma$ with probability $0.5$, where $\sigma$ is the standard deviation of $N_{nm}$, which we estimate by a simple algorithm described in the Appendix. Using this method, the required regularization parameter can be found by minimizing the function
\be\label{fld}
S(\lambda) = \left\|P\;\text{DFT}^{-1}\left[\frac{\lambda |\widehat{\nabla}^2_{kl}|^2(\widehat{\widetilde B}_{kl})}{|\widehat{K}_{1, kl}|^2+|\widehat{K}_{2, kl}|^2+\lambda |\widehat{\nabla}^2_{kl}|^2}\right]\right\|^2_2-2\widehat C(\lambda),
\ee
where
\be\label{fld2}
\widehat C(\lambda) = \|PN_1\|^2_2-\Re \left[\left\langle PN_1, P \; \text{DFT}^{-1}\left[ H_{kl}(\lambda)\widehat{N}_{1,kl}\right]\right\rangle_2\right].
\ee
The projected SURE for GI-SURE scheme is obtained from \eqref{fld}-\eqref{fld2} by replacing $|\widehat{K}_{1, kl}|^2+|\widehat{K}_{2, kl}|^2$ in \eqref{fld} with $|\widehat{K}_{kl}|^2$.

Thus, to apply the DI method the following steps have to be taken:
\begin{enumerate}
  \item Compute the extended field $\widetilde B_z$ using \eqref{eq:flipsForData} (if currents cross the image boundary).
  \item Compute $\lambda$ by minimizing either \eqref{G222} for DI-GCV or \eqref{fld} for DI-SURE.
  \item Obtain the currents using \eqref{2convRegSoln_a}-\eqref{2convRegSoln_b}, with $\widetilde B_z$ substituted for $B_z$.
  \item Take only the currents lying inside the measurement window.
\end{enumerate}
In Sec. \ref{Sec2} we presented the algorithm for implementation of GI without discussing the possibility of symmetric extension of the field. Implementation of the GI method with the symmetric extension is similar to the DI method presented in the current Section in the sense that the calculations are performed using a symmetrically extended data and the result is taken from inside the measurement window. In the next Section we compare the performance of these two methods through several numerical examples.
\section{Numerical results \label{Numres}}

In the present section we apply the above-proposed inversion algorithms to three examples of current distributions in thin films. Each example consists of a square sample with side length $l_1$ and a square hole in the center with side length $l_2$. The circulating currents in the samples are determined by numerically solving the Ginzburg-Landau equations in the presence of an applied external magnetic field \cite{TDGL1}. The measured window, however, may contain only part of the sample and is further corrupted by noise,
\be
B_{nm} = B_{true,nm}+\sigma W_{nm}.
\ee
Here $B_{true,nm}$ is calculated using the Biot-Savart law taking into account the currents that flow in the entire sample and $\sigma W_{nm}$ is Gaussian white noise with standard deviation $\sigma=s\max{|B_{true,nm}|}$ and $s \in \{10^{-1}, 10^{-2}, 10^{-3}\}$. Below, the magnetic field is given in units of Gauss, the electric current in $mA$ and the length in $\mu \text{m}$ so that $\mu_0=4\pi$. We set the grid step to $\Delta x=\Delta y=1 ~\mu \text{m}$, the imaging height to $h=1 ~\mu \text{m}$ and the thickness of the sample to $d=0.05~\mu m$.

\begin{figure}
    \centering
    \includegraphics[width=\textwidth]{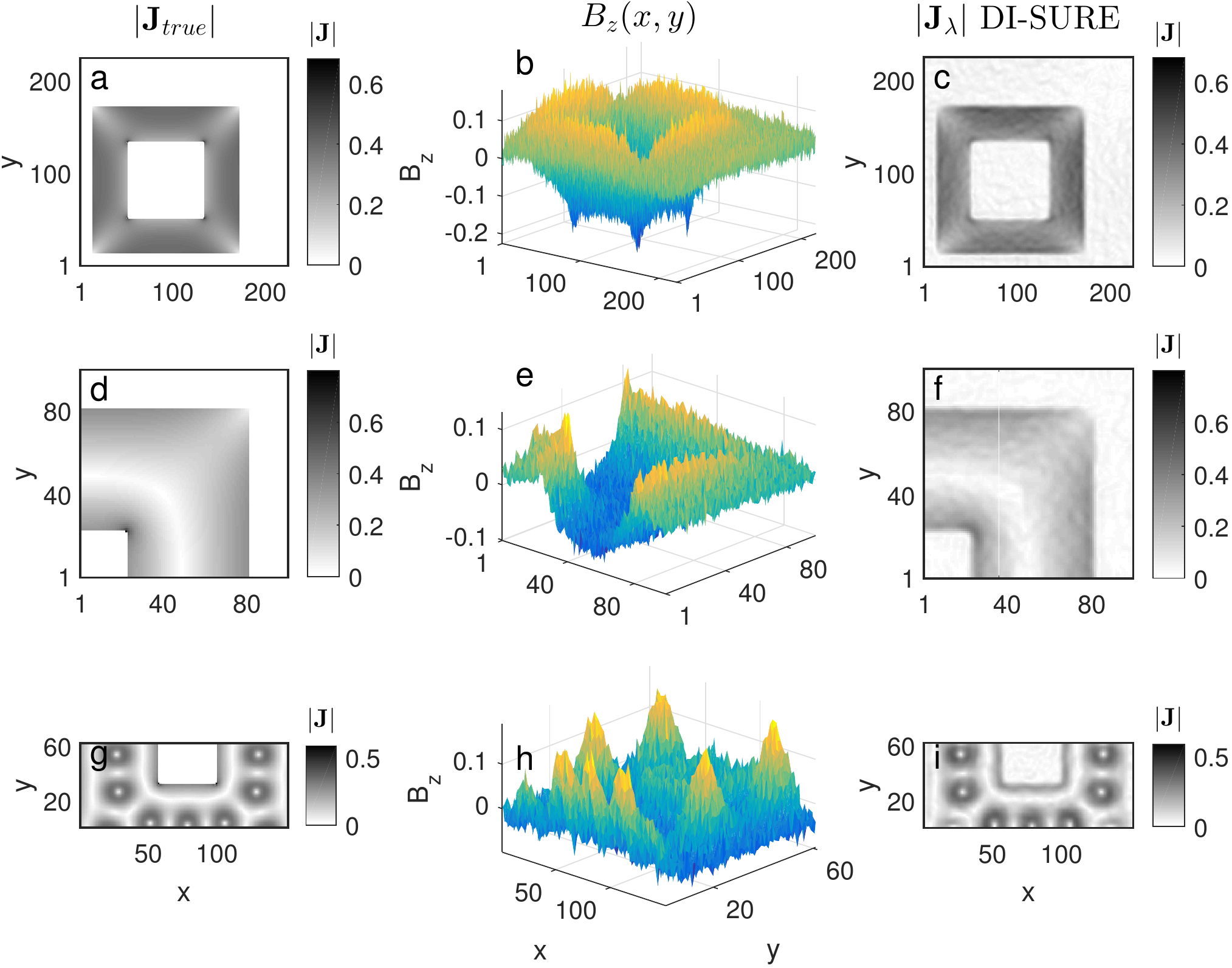}
    \caption{(Left column) The true current density in samples A (top row), B (middle row), and C (bottom row). (Middle column) The corresponding calculated magnetic field at height $h=1$ $\mu$m perturbed by a noise of $s=10^{-1}$. (Right column) The current density reconstructed from the noisy \(B_z(x,y)\) using the DI-SURE scheme. The edges of the plots coincide with the measurement window. The $x$ and $y$ axes are in units of $\mu$m, the current density in mA/$\mu$m$^2$, and the field in Gauss.}\label{ex3}
\end{figure}

Sample A consists of a loop of outer and inner side lengths $l_1=161 ~\mu \text{m}$ and $l_2=83~\mu \text{m}$, respectively, and a clockwise current flow (see Fig. \ref{ex3}(a)). The sample is entirely contained in the square measurement window of side length $223 ~\mu \text{m}$, making this example solvable without the reflection procedure.

Sample B consists of a square loop with $l_1=159~\mu \text{m}$ and $l_2=43~\mu \text{m}$, that carries a counter-clockwise current flow in the inner part of the loop and a clockwise flow in the outer part, as shown in Fig. \ref{ex3}(d). The square measurement window of side length $99 ~\mu \text{m}$ includes one corner of the loop only, making the use of reflection necessary for accurate current reconstruction.

The third example, sample C, shown in Fig. \ref{ex3}(g), consists of a square loop with $l_1=161 ~\mu \text{m}$ and $l_2=41~\mu \text{m}$. The loop has several vortices distributed in the sample. The measurement window of side length $151 ~\mu \text{m}$ by $61~\mu \text{m}$ cuts the loop from all four sides. The bottom cut is very close to the cores of the vortices, representing a challenging inversion problem that can be dealt with by our reflection rule as demonstrated below.

The true current densities in samples A, B and C are shown in Figs. \ref{ex3}(a), \ref{ex3}(d) and \ref{ex3}(g). Magnetic fields generated by the currents are corrupted by noise with $s=10^{-1}$, and are shown in the central column (Figs. \ref{ex3}(b), \ref{ex3}(e), and \ref{ex3}(h)). In the right column (Figs. \ref{ex3}(c), \ref{ex3}(f), and \ref{ex3}(i)) we present the current densities reconstructed from the corresponding magnetic fields in the central column, using the DI-SURE scheme. Comparing between the left and the right columns we can conclude that the quality of the reconstruction is high, notwithstanding the high noise level in the central column. The success of current reconstruction in sample C is particularly impressive in view of the presence of vortices that are cut through by the measurement window, the reconstruction of which can be assumed to require more sophisticated boundary conditions.

\begin{figure}
 \centering
        \includegraphics[trim = .1 .1 .1 .1, clip, width=\textwidth]{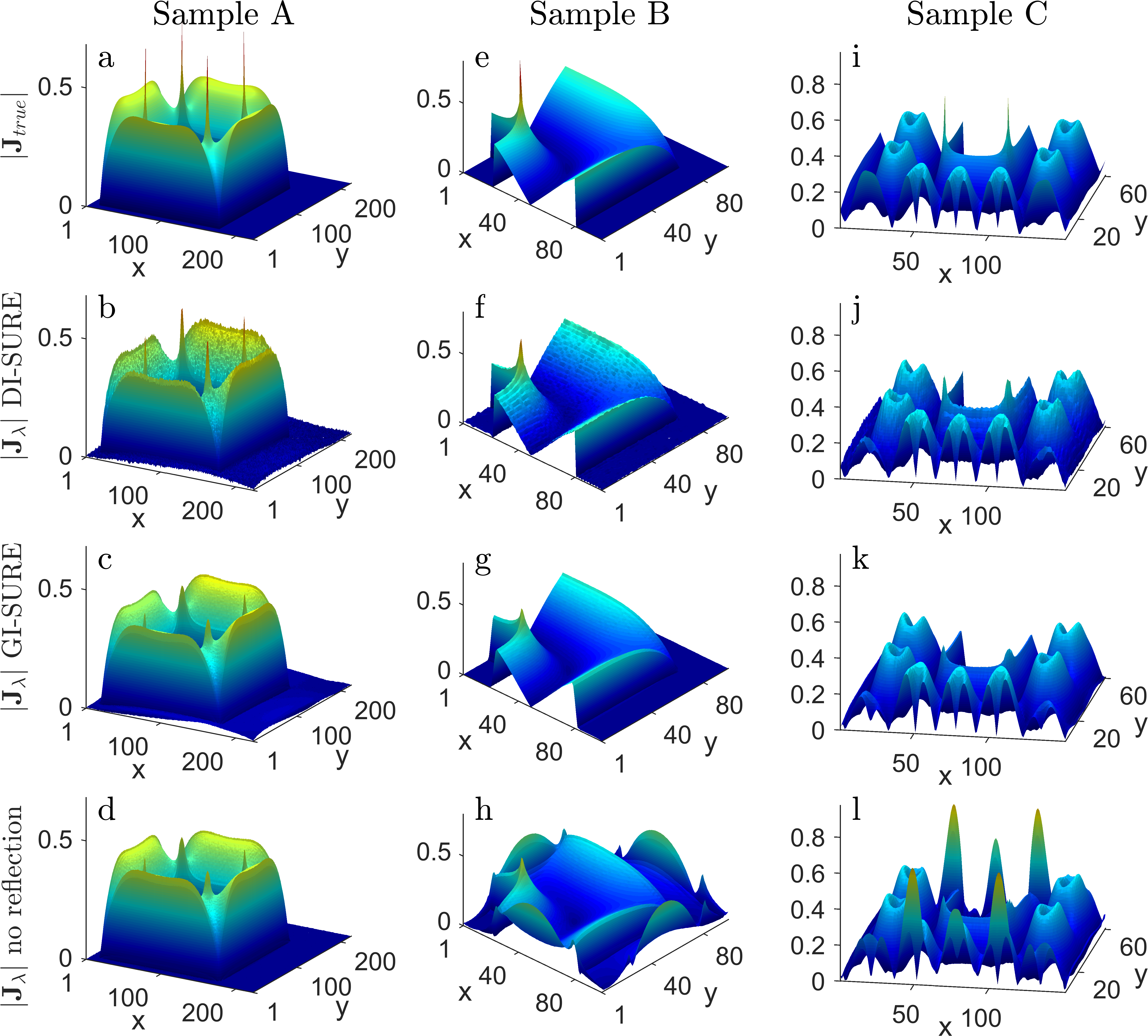}
      \caption{(Top row) Surfaces of the true current density. (Second row) The current density reconstructed with DI-SURE using the symmetric extension of the field. (Third row) The current density reconstructed with GI-SURE using the symmetric extension of the field. (Bottom row) The current density reconstructed with GI-SURE without using the symmetric extension of the field. The results are presented for $s=10^{-3}$ in samples A (left column), B (middle column), and C (right column). The $x$ and $y$ axes are in units of $\mu$m and the current density in mA/$\mu$m$^2$.}\label{fig:RecSurf}
\end{figure}

In order to highlight the difference between the GI and the DI methods, we present in Fig. \ref{fig:RecSurf} the results of current reconstruction in samples A, B and C in the presence of a low noise with $s=10^{-3}$. The true current densities for these three samples are shown in Figs. \ref{fig:RecSurf}(a), \ref{fig:RecSurf}(e) and \ref{fig:RecSurf}(i) and their reconstructions using prior symmetric extension of the magnetic field by either the DI-SURE or the GI-SURE method are shown in Figs. \ref{fig:RecSurf}(b), \ref{fig:RecSurf}(f) and \ref{fig:RecSurf}(j) and Figs. \ref{fig:RecSurf}(c), \ref{fig:RecSurf}(g) and \ref{fig:RecSurf}(k), respectively. Comparing these results we observe that the GI method produces a smoother solution, while the DI method provides a better reconstruction of the sharp features in the current density. The reason for this, as mentioned above, is the strong smoothing by the GI method, which uses two filters for current reconstruction. To emphasize the advantage of the reflection procedure we present in the last row of Fig. \ref{fig:RecSurf} ((d), (h) and (l)) the results of reconstructions using GI-SURE without prior symmetric extension of the magnetic field. As expected, the reconstruction remains accurate for sample A, where the currents are closed within the measurement window. However, for samples B and C the reconstruction is highly inaccurate, especially near the window boundary.

\begin{figure}
        \centering
        \includegraphics[width=\textwidth]{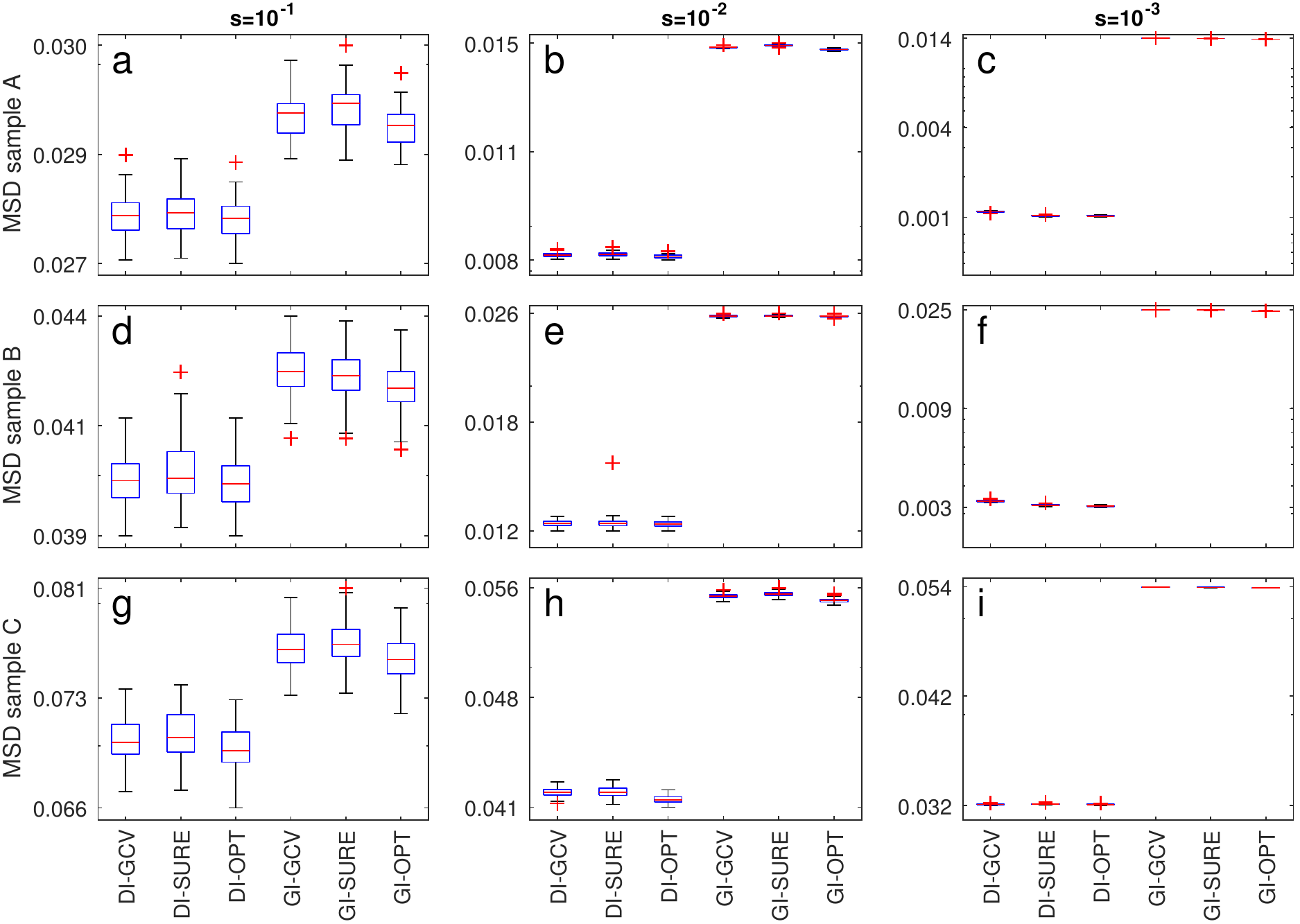}
    \caption{Boxplots of the MSDs of the currents reconstructed in samples A (top row), B (middle row), and C (bottom row) for noise levels of $s=10^{-1}$ (left column), $s=10^{-2}$ (middle column), and $s=10^{-3}$ (right column). The MSD values for solutions obtained using the optimal values of $\lambda$ are denoted DI-OPT and GI-OPT.}\label{fig:Boxplots}
\end{figure}

We now perform a quantitative analysis to compare the results of the four presented inversion methods by comparing their MSDs, which are defined in \eqref{MSD}. We apply the inversion procedure to each sample 100 times, each time using a different noise realization, and present boxplots of the MSD values in Fig. \ref{fig:Boxplots}. The boxplots graphically depict the results by splitting them into quartiles, so that each box spans the range that contains the second and third quartiles, termed the interquartile range (i.e., the middle 50\% of the data). The horizontal line in each box denotes the median, while the error bars span 150\% of the interquartile range above the third quartile and below the second quartile. Any point outside this interval  is denoted by '+' and considered an outlier.

The MSD of the reconstructions in Fig. \ref{fig:Boxplots} is given alongside the best possible MSD, which is calculated using the $\lambda$ that minimizes the MSD function \eqref{MSD}. The accuracy of the reconstruction in sample A, where the current does not cross the image boundary and therefore the symmetric extension is not performed, is shown in Fig. \ref{fig:Boxplots}(a)-\ref{fig:Boxplots}(c).  In contrast, in samples B and C the current crosses the measurement boundary and therefore a symmetric extension of the measured magnetic field is necessary. The accuracy of the reconstruction of these samples is shown in Figs. \ref{fig:Boxplots}(d)-\ref{fig:Boxplots}(f) and \ref{fig:Boxplots}(g)-\ref{fig:Boxplots}(i). As can be seen, both methods are close to their optimum solutions, but the DI methods consistently achieve a lower MSD compared to the GI methods, in all examples. The advantage of the DI method becomes more significant at conditions of lower noise since, in contrast to the GI method that uses the Savitsky-Golay filtering upon differentiating the $g$ function, the DI method does not use an additional filter and thus preserves the sharp features of the solution. This effect is particularly pronounced in Fig. \ref{fig:Boxplots}(c) where the DI methods provide an MSD that is more than an order of magnitude lower.

\begin{figure}
        \centering
        \includegraphics[width=\textwidth]{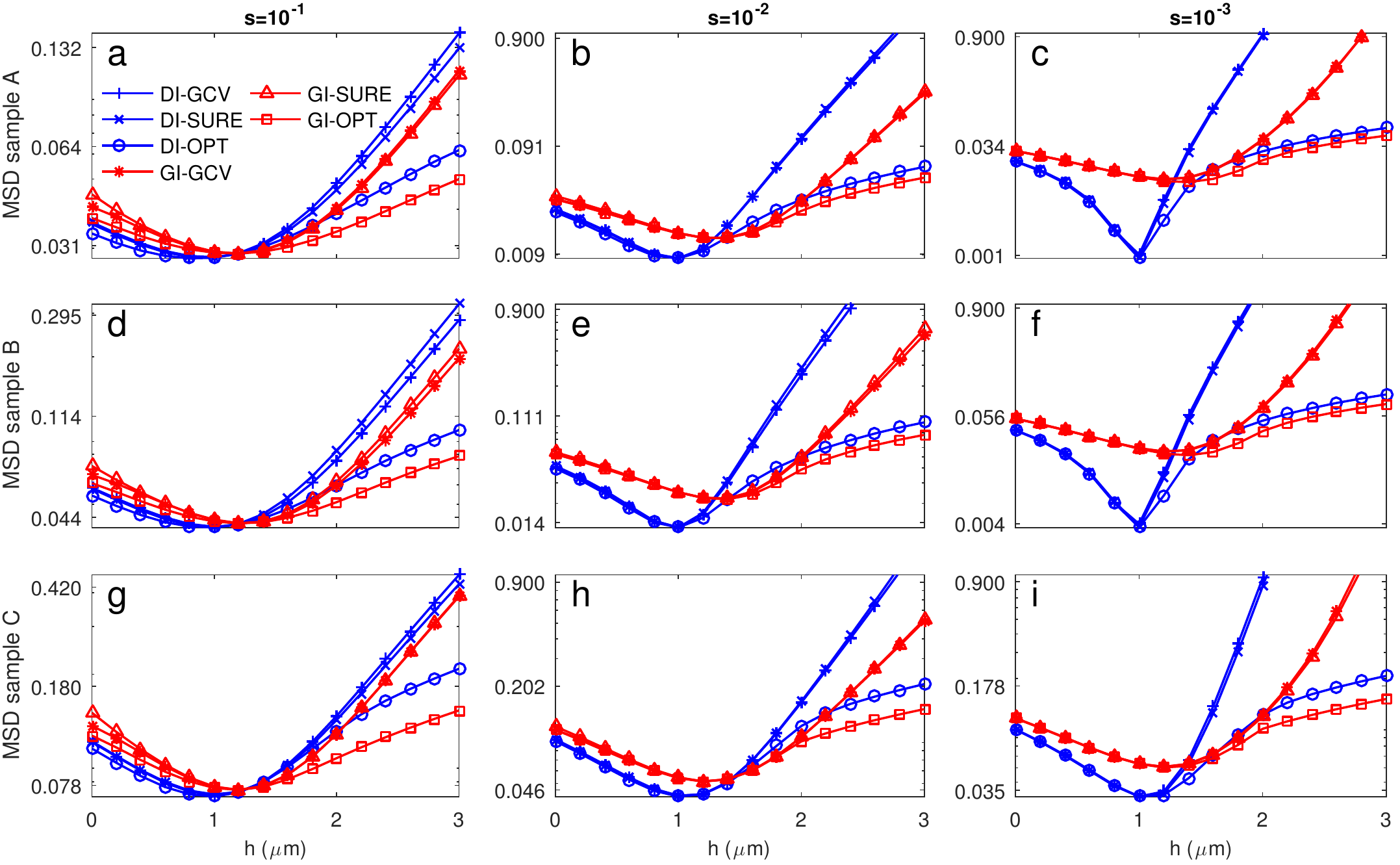}
    \caption{Plots of the MSDs of current reconstructions as a function of the assumed height $h$, given the true height of $h_{true}=1$ $\mu$m in samples A (top row), B (middle row), and C (bottom row) for noise levels of $s=10^{-1}$ (left column), $s=10^{-2}$ (middle column), and $s=10^{-3}$ (right column). The MSD values for solutions obtained using the optimal values of $\lambda$ are denoted DI-OPT and GI-OPT.}
\label{fig:estH}
\end{figure}

Next, we use the magnetic field simulated at $h_{true}=1~\mu \text{m}$ and assume, as in real measurements, that the exact values of the true height are not known. The accuracy of reconstructions of the currents, assuming different heights $h$, is shown in Fig. \ref{fig:estH}. The MSD curves in the presence of noise are not steep around the true height, and therefore the reconstruction remains reliable even with a wrong estimation of $h_{true}$, with underestimation preferable to overestimation. The DI methods provides the lowest MSD value at $h_{true}$ while the lowest MSD values using the GI method are obtained for values slightly above the true height. For heights comparable to and lower than $h_{true}$ the DI method provides consistently lower MSD values, while for values above $h_{true}$ GI attains lower MSD values.

\begin{figure}
        \centering
        \includegraphics[width=\textwidth]{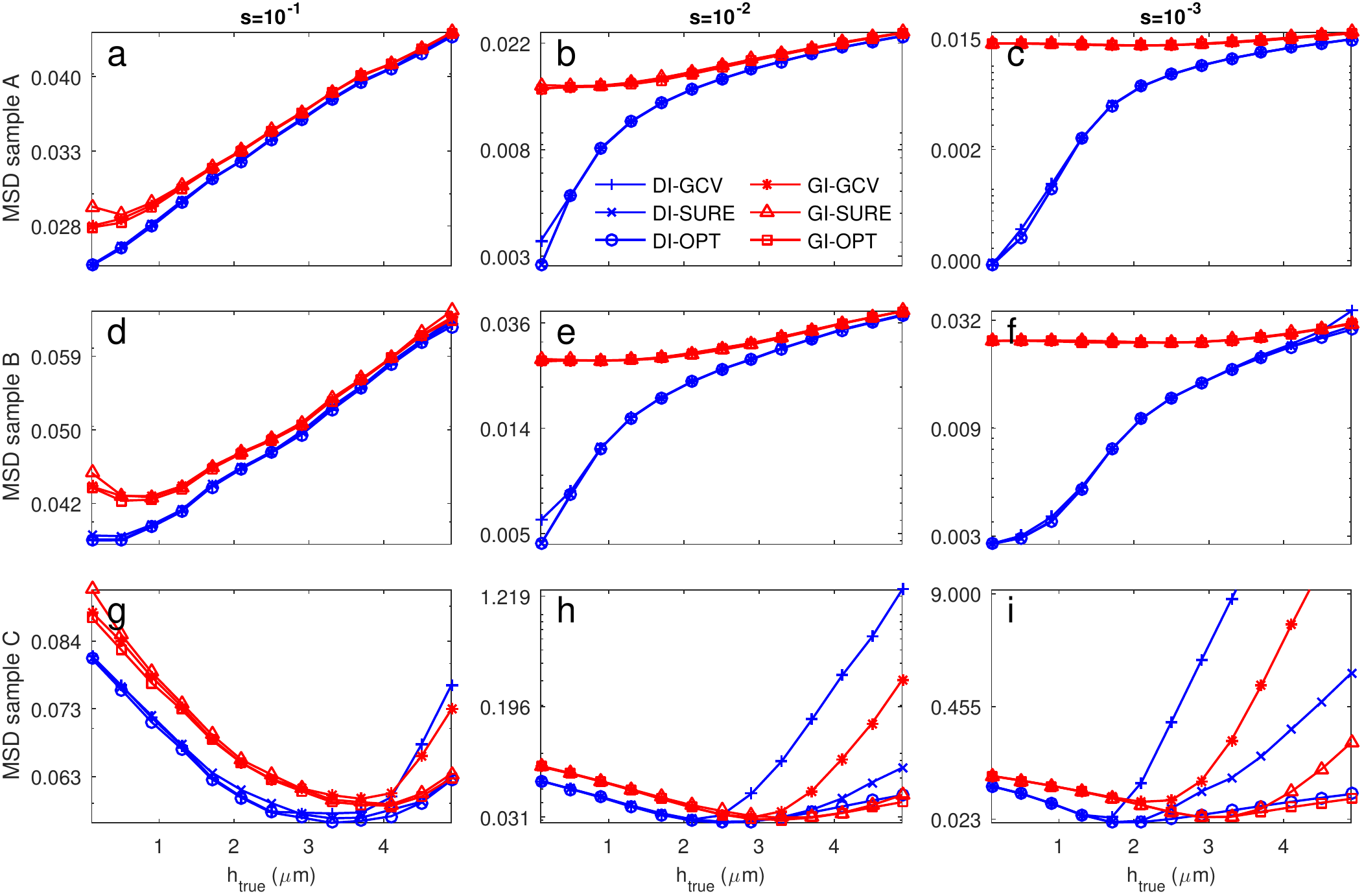}
    \caption{Plots of the MSDs of current reconstructions as a function of the true height $h_{true}$ in samples A (top row), B (middle row), and C (bottom row) for noise levels of $s=10^{-1}$ (left column), $s=10^{-2}$ (middle column), and $s=10^{-3}$ (right column). The MSD values for solutions obtained using the optimal values of $\lambda$ are denoted DI-OPT and GI-OPT.} \label{fig:varHt}
\end{figure}

Finally, we analyze the accuracy of the inversion methods at different measurement heights, assuming that the reconstruction is performed at a true, variable height. In Fig. \ref{fig:varHt} we show the MSD curves of the reconstructed currents as a function of the true height $h_{true}$. At low heights the DI methods result in lower MSD, but at large heights the GI methods may have an advantage in some cases, as exemplified by sample C, in which highly irregular currents cross the measurement window. Another observation is that when the currents are not closed within the measurement window the GCV regularization becomes ineffective for both the DI and the GI methods at larger heights, while the projected SURE regularization remains accurate for a much wider range of measurement heights.

\section{Conclusions}
Reconstruction of nanoscale electric current distributions in thin samples is important for the characterization of new low-dimensional materials and the evaluation of electric devices. A general scheme for the reconstruction of electric currents from a measured out-of-plane component of the magnetic field above the sample surface was presented. Our approach comprises three innovative parts: (1) a direct formulation of the inversion problem,(2) a symmetric extension of the measured magnetic field, and (3) an enhanced method for determination of the regularization parameter. Using the method of Feldmann \cite{Feldman} as reference, we show that direct formulation of the reconstruction problem allows us to improve the accuracy of current reconstruction, especially at regions that contain sharp alterations, while the symmetric extension of the measured magnetic field enables reconstruction of non-closed currents. Finally, our scheme for determination of the regularization parameter makes current reconstruction possible over an extended range of measurement heights.

We presented here several methods for current reconstruction. Two of the methods, the DI-GCV and the DI-SURE, reconstruct current fields directly, while two other methods, the GI-GCV and the GI-SURE, reconstruct the stream function from which the current fields are obtained using a smoothed differentiation. The advantage of the DI methods is most pronounced at smaller heights, which are preferable for resolution of a local magnetic structure. At small heights (relative to the grid spacing) the DI methods indeed demonstrate a significant gain in accuracy in all our numerical experiments. In the presence of large noise and at large heights the DI schemes still outperform the GI schemes for samples with closed currents, albeit with a smaller gain in accuracy, with the GI schemes providing the extra benefit of a smoother solution. In our numerical simulations the difference between GCV-based and SURE-based methods is very small. Therefore, for closed currents we suggest using the DI-GCV at all times, unless a smooth solution is required and an accurate reconstruction of sharp changes of currents is not needed, in which case we suggest using the GI-GCV.

When currents cross the measurement boundary, a symmetric extension of the measured field is used to effectively approximate them and enable usage of the DFT, which requires periodicity. For small heights DI-GCV and DI-SURE schemes have a very similar accuracy, which is superior to that of GI schemes, similarly to the case of closed currents. For large heights however GCV-based schemes become less accurate due to the poor reconstruction close to the measurement boundary, rendering the SURE-based schemes essential. In this case the use of the GI-SURE method is recommended.

\begin{acknowledgments}
This research was supported by the US-Israel Binational Science Foundation (BSF grant 2014155), by the Israel Science Foundation (grant No. 132/14), and by Rosa and Emilio Segr\'{e} Research Award.
\end{acknowledgments}

\appendix
\section{Method for variance estimation}
Accurate estimation of the variance $\sigma^2$ of the measured magnetic field $B_{nm}$ is important for determination of the regularization parameter using the projected SURE method in Sec. \ref{RegParSec}. Here we describe an algorithm, which is based on the ideas developed in \cite{PP2016} and carried over to the discrete Fourier space using arguments given in \cite{HansenFFT}. Assuming the image $B_{nm}$ represents a smooth magnetic field corrupted by noise, the Fourier coefficients $\widehat{B}_{nm}$ can be divided into two parts, the part containing the information about the magnetic field, and the other part containing the noise. We shift $\widehat{B}_{nm}$ such that the zero Fourier mode is located at $n=m=1$, while the high frequency noise coefficients occupy the center of $\widehat{B}_{nm}$. For a successful estimation it is sufficient to find the region in the frequency space $\widehat{B}_{nm}$ which contains only noise and compute its sample variance. For this purpose we construct a nested sequence of rectangular sub-matrices $M_{j,nm}$ such that $M_{1,nm}$ equals the entire matrix $\widehat{B}_{nm}$ and $M_{q,nm}$ contains only the highest Fourier modes in the center of $\widehat{B}_{nm}$. We then define the function
\[
V(j)=\sum_{nm} |M_{j,nm}|^2/N_j,
\]
where $N_j$ is the number of elements in $M_{j,nm}$. It is shown in \cite{PP2016} that the values of the curve $V(j)$, which approximates the expected value of $|M_{k,nm}|^2$, decreases and levels off at $\sigma^2$. We thus find an index $k_0$ in the flat region of $V(k)$ by minimizing $\|\log(V(k+1))-\log(V(k))\|$ and obtain our estimate as $\sigma^2\simeq V(k_0)$.

\bibliographystyle{apsrev4-1v1}
\bibliography{WIP}

\begin{thebibliography}{73}%
\makeatletter
\providecommand \@ifxundefined [1]{%
 \@ifx{#1\undefined}
}%
\providecommand \@ifnum [1]{%
 \ifnum #1\expandafter \@firstoftwo
 \else \expandafter \@secondoftwo
 \fi
}%
\providecommand \@ifx [1]{%
 \ifx #1\expandafter \@firstoftwo
 \else \expandafter \@secondoftwo
 \fi
}%
\providecommand \natexlab [1]{#1}%
\providecommand \enquote  [1]{``#1''}%
\providecommand \bibnamefont  [1]{#1}%
\providecommand \bibfnamefont [1]{#1}%
\providecommand \citenamefont [1]{#1}%
\providecommand \href@noop [0]{\@secondoftwo}%
\providecommand \href [0]{\begingroup \@sanitize@url \@href}%
\providecommand \@href[1]{\@@startlink{#1}\@@href}%
\providecommand \@@href[1]{\endgroup#1\@@endlink}%
\providecommand \@sanitize@url [0]{\catcode `\\12\catcode `\$12\catcode
  `\&12\catcode `\#12\catcode `\^12\catcode `\_12\catcode `\%12\relax}%
\providecommand \@@startlink[1]{}%
\providecommand \@@endlink[0]{}%
\providecommand \url  [0]{\begingroup\@sanitize@url \@url }%
\providecommand \@url [1]{\endgroup\@href {#1}{\urlprefix }}%
\providecommand \urlprefix  [0]{URL }%
\providecommand \Eprint [0]{\href }%
\providecommand \doibase [0]{http://dx.doi.org/}%
\providecommand \selectlanguage [0]{\@gobble}%
\providecommand \bibinfo  [0]{\@secondoftwo}%
\providecommand \bibfield  [0]{\@secondoftwo}%
\providecommand \translation [1]{[#1]}%
\providecommand \BibitemOpen [0]{}%
\providecommand \bibitemStop [0]{}%
\providecommand \bibitemNoStop [0]{.\EOS\space}%
\providecommand \EOS [0]{\spacefactor3000\relax}%
\providecommand \BibitemShut  [1]{\csname bibitem#1\endcsname}%
\let\auto@bib@innerbib\@empty
\bibitem [{\citenamefont {Dinner}\ \emph {et~al.}(2007)\citenamefont {Dinner},
  \citenamefont {Moler}, \citenamefont {Feldmann},\ and\ \citenamefont
  {Beasley}}]{Feldman2}%
  \BibitemOpen
  \bibfield  {author} {\bibinfo {author} {\bibfnamefont {R.~B.}\ \bibnamefont
  {Dinner}}, \bibinfo {author} {\bibfnamefont {K.~A.}\ \bibnamefont {Moler}},
  \bibinfo {author} {\bibfnamefont {D.~M.}\ \bibnamefont {Feldmann}}, \ and\
  \bibinfo {author} {\bibfnamefont {M.~R.}\ \bibnamefont {Beasley}},\ }\bibinfo
  {title} {{Imaging ac losses in superconducting films via scanning Hall probe
  microscopy}},\ \href {\doibase 10.1103/PhysRevB.75.144503} {\bibfield
  {journal} {\bibinfo  {journal} {Phys. Rev. B}\ }\textbf {\bibinfo {volume}
  {75}},\ \bibinfo {pages} {144503} (\bibinfo {year} {2007})}\BibitemShut
  {NoStop}%
\bibitem [{\citenamefont {Jooss}\ \emph {et~al.}(2002)\citenamefont {Jooss},
  \citenamefont {Albrecht},\ and\ \citenamefont {Kuhn}}]{Jooss1}%
  \BibitemOpen
  \bibfield  {author} {\bibinfo {author} {\bibfnamefont {C.}~\bibnamefont
  {Jooss}}, \bibinfo {author} {\bibfnamefont {J.}~\bibnamefont {Albrecht}}, \
  and\ \bibinfo {author} {\bibfnamefont {H.}~\bibnamefont {Kuhn}},\ }\bibinfo
  {title} {{Magneto-optical studies of current distributions in high-T$_c$
  superconductors}},\ \href {\doibase 10.1088/0034-4885/65/5/202} {\bibfield
  {journal} {\bibinfo  {journal} {Reports Prog. Phys.}\ }\textbf {\bibinfo
  {volume} {65}},\ \bibinfo {pages} {651} (\bibinfo {year} {2002})}\BibitemShut
  {NoStop}%
\bibitem [{\citenamefont {Jooss}\ \emph {et~al.}(1998)\citenamefont {Jooss},
  \citenamefont {Warthmann}, \citenamefont {Forkl},\ and\ \citenamefont
  {Kronm{\"{u}}ller}}]{JoossN}%
  \BibitemOpen
  \bibfield  {author} {\bibinfo {author} {\bibfnamefont {C.}~\bibnamefont
  {Jooss}}, \bibinfo {author} {\bibfnamefont {R.}~\bibnamefont {Warthmann}},
  \bibinfo {author} {\bibfnamefont {A.}~\bibnamefont {Forkl}}, \ and\ \bibinfo
  {author} {\bibfnamefont {H.}~\bibnamefont {Kronm{\"{u}}ller}},\ }\bibinfo
  {title} {{High-resolution magneto-optical imaging of critical currents in
  {YB}a$_2${C}u$_3${O}$_{7-\delta}$ thin films}},\ \href {\doibase
  10.1016/S0921-4534(97)01887-X} {\bibfield  {journal} {\bibinfo  {journal}
  {Phys. C Supercond.}\ }\textbf {\bibinfo {volume} {299}},\ \bibinfo {pages}
  {215} (\bibinfo {year} {1998})}\BibitemShut {NoStop}%
\bibitem [{\citenamefont {Pashitski}\ \emph {et~al.}(1997)\citenamefont
  {Pashitski}, \citenamefont {Gurevich}, \citenamefont {Polyanskii},
  \citenamefont {Larbalestier}, \citenamefont {Goyal}, \citenamefont {Specht},
  \citenamefont {Kroeger}, \citenamefont {DeLuca},\ and\ \citenamefont
  {Tkaczyk}}]{JRecPashitski17011997}%
  \BibitemOpen
  \bibfield  {author} {\bibinfo {author} {\bibfnamefont {A.~E.}\ \bibnamefont
  {Pashitski}}, \bibinfo {author} {\bibfnamefont {A.}~\bibnamefont {Gurevich}},
  \bibinfo {author} {\bibfnamefont {A.~A.}\ \bibnamefont {Polyanskii}},
  \bibinfo {author} {\bibfnamefont {D.~C.}\ \bibnamefont {Larbalestier}},
  \bibinfo {author} {\bibfnamefont {A.}~\bibnamefont {Goyal}}, \bibinfo
  {author} {\bibfnamefont {E.~D.}\ \bibnamefont {Specht}}, \bibinfo {author}
  {\bibfnamefont {D.~M.}\ \bibnamefont {Kroeger}}, \bibinfo {author}
  {\bibfnamefont {J.~A.}\ \bibnamefont {DeLuca}}, \ and\ \bibinfo {author}
  {\bibfnamefont {J.~E.}\ \bibnamefont {Tkaczyk}},\ }\bibinfo {title}
  {{Reconstruction of current flow and imaging of current-limiting defects in
  polycrystalline superconducting films}},\ \href {\doibase
  10.1126/science.275.5298.367} {\bibfield  {journal} {\bibinfo  {journal}
  {Science}\ }\textbf {\bibinfo {volume} {275}},\ \bibinfo {pages} {367}
  (\bibinfo {year} {1997})}\BibitemShut {NoStop}%
\bibitem [{\citenamefont {Carrera}\ \emph {et~al.}(2011)\citenamefont
  {Carrera}, \citenamefont {Granados}, \citenamefont {Amor{\'{o}}s},
  \citenamefont {Maynou}, \citenamefont {Puig},\ and\ \citenamefont
  {Obradors}}]{HTM1}%
  \BibitemOpen
  \bibfield  {author} {\bibinfo {author} {\bibfnamefont {M.}~\bibnamefont
  {Carrera}}, \bibinfo {author} {\bibfnamefont {X.}~\bibnamefont {Granados}},
  \bibinfo {author} {\bibfnamefont {J.}~\bibnamefont {Amor{\'{o}}s}}, \bibinfo
  {author} {\bibfnamefont {R.}~\bibnamefont {Maynou}}, \bibinfo {author}
  {\bibfnamefont {T.}~\bibnamefont {Puig}}, \ and\ \bibinfo {author}
  {\bibfnamefont {X.}~\bibnamefont {Obradors}},\ }\bibinfo {title} {{Computed
  current distribution in HTS tapes obtained from Hall magnetic mapping by
  inverse problem solution}},\ \href {\doibase 10.1109/TASC.2010.2089596}
  {\bibfield  {journal} {\bibinfo  {journal} {IEEE Trans. Appl. Supercond.}\
  }\textbf {\bibinfo {volume} {21}},\ \bibinfo {pages} {9133} (\bibinfo {year}
  {2011})}\BibitemShut {NoStop}%
\bibitem [{\citenamefont {Sun}\ \emph {et~al.}(2014)\citenamefont {Sun},
  \citenamefont {Tsuchiya}, \citenamefont {Taen}, \citenamefont {Yamada},
  \citenamefont {Pyon}, \citenamefont {Sugimoto}, \citenamefont {Ekino},
  \citenamefont {Shi},\ and\ \citenamefont {Tamegai}}]{Sun2014}%
  \BibitemOpen
  \bibfield  {author} {\bibinfo {author} {\bibfnamefont {Y.}~\bibnamefont
  {Sun}}, \bibinfo {author} {\bibfnamefont {Y.}~\bibnamefont {Tsuchiya}},
  \bibinfo {author} {\bibfnamefont {T.}~\bibnamefont {Taen}}, \bibinfo {author}
  {\bibfnamefont {T.}~\bibnamefont {Yamada}}, \bibinfo {author} {\bibfnamefont
  {S.}~\bibnamefont {Pyon}}, \bibinfo {author} {\bibfnamefont {A.}~\bibnamefont
  {Sugimoto}}, \bibinfo {author} {\bibfnamefont {T.}~\bibnamefont {Ekino}},
  \bibinfo {author} {\bibfnamefont {Z.}~\bibnamefont {Shi}}, \ and\ \bibinfo
  {author} {\bibfnamefont {T.}~\bibnamefont {Tamegai}},\ }\bibinfo {title}
  {{Dynamics and mechanism of oxygen annealing in
  {F}e$_{1+y}${T}e$_{0.6}${S}e$_{0.4}$ single crystal}},\ \href {\doibase
  10.1038/srep04585} {\bibfield  {journal} {\bibinfo  {journal} {Sci. Rep.}\
  }\textbf {\bibinfo {volume} {4}},\ \bibinfo {pages} {4585} (\bibinfo {year}
  {2014})}\BibitemShut {NoStop}%
\bibitem [{\citenamefont {Nowack}\ \emph {et~al.}(2013)\citenamefont {Nowack},
  \citenamefont {Spanton}, \citenamefont {Baenninger}, \citenamefont
  {K{\"{o}}nig}, \citenamefont {Kirtley}, \citenamefont {Kalisky},
  \citenamefont {Ames}, \citenamefont {Leubner}, \citenamefont {Br{\"{u}}ne},
  \citenamefont {Buhmann}, \citenamefont {Molenkamp}, \citenamefont
  {Goldhaber-Gordon},\ and\ \citenamefont {Moler}}]{SCANNING61}%
  \BibitemOpen
  \bibfield  {author} {\bibinfo {author} {\bibfnamefont {K.~C.}\ \bibnamefont
  {Nowack}}, \bibinfo {author} {\bibfnamefont {E.~M.}\ \bibnamefont {Spanton}},
  \bibinfo {author} {\bibfnamefont {M.}~\bibnamefont {Baenninger}}, \bibinfo
  {author} {\bibfnamefont {M.}~\bibnamefont {K{\"{o}}nig}}, \bibinfo {author}
  {\bibfnamefont {J.~R.}\ \bibnamefont {Kirtley}}, \bibinfo {author}
  {\bibfnamefont {B.}~\bibnamefont {Kalisky}}, \bibinfo {author} {\bibfnamefont
  {C.}~\bibnamefont {Ames}}, \bibinfo {author} {\bibfnamefont {P.}~\bibnamefont
  {Leubner}}, \bibinfo {author} {\bibfnamefont {C.}~\bibnamefont
  {Br{\"{u}}ne}}, \bibinfo {author} {\bibfnamefont {H.}~\bibnamefont
  {Buhmann}}, \bibinfo {author} {\bibfnamefont {L.~W.}\ \bibnamefont
  {Molenkamp}}, \bibinfo {author} {\bibfnamefont {D.}~\bibnamefont
  {Goldhaber-Gordon}}, \ and\ \bibinfo {author} {\bibfnamefont {K.~A.}\
  \bibnamefont {Moler}},\ }\bibinfo {title} {{Imaging currents in HgTe quantum
  wells in the quantum spin Hall regime}},\ \href {\doibase 10.1038/nmat3682}
  {\bibfield  {journal} {\bibinfo  {journal} {Nat. Mater.}\ }\textbf {\bibinfo
  {volume} {12}},\ \bibinfo {pages} {787} (\bibinfo {year} {2013})}\BibitemShut
  {NoStop}%
\bibitem [{\citenamefont {Spanton}\ \emph {et~al.}(2014)\citenamefont
  {Spanton}, \citenamefont {Nowack}, \citenamefont {Du}, \citenamefont
  {Sullivan}, \citenamefont {Du},\ and\ \citenamefont {Moler}}]{Spanton2014}%
  \BibitemOpen
  \bibfield  {author} {\bibinfo {author} {\bibfnamefont {E.~M.}\ \bibnamefont
  {Spanton}}, \bibinfo {author} {\bibfnamefont {K.~C.}\ \bibnamefont {Nowack}},
  \bibinfo {author} {\bibfnamefont {L.}~\bibnamefont {Du}}, \bibinfo {author}
  {\bibfnamefont {G.}~\bibnamefont {Sullivan}}, \bibinfo {author}
  {\bibfnamefont {R.-R.}\ \bibnamefont {Du}}, \ and\ \bibinfo {author}
  {\bibfnamefont {K.~A.}\ \bibnamefont {Moler}},\ }\bibinfo {title} {{Images of
  edge current in InAs/GaSb quantum wells}},\ \href {\doibase
  10.1103/PhysRevLett.113.026804} {\bibfield  {journal} {\bibinfo  {journal}
  {Phys. Rev. Lett.}\ }\textbf {\bibinfo {volume} {113}},\ \bibinfo {pages}
  {026804} (\bibinfo {year} {2014})}\BibitemShut {NoStop}%
\bibitem [{\citenamefont {Kalisky}\ \emph {et~al.}(2013)\citenamefont
  {Kalisky}, \citenamefont {Spanton}, \citenamefont {Noad}, \citenamefont
  {Kirtley}, \citenamefont {Nowack}, \citenamefont {Bell}, \citenamefont
  {Sato}, \citenamefont {Hosoda}, \citenamefont {Xie}, \citenamefont {Hikita},
  \citenamefont {Woltmann}, \citenamefont {Pfanzelt}, \citenamefont {Jany},
  \citenamefont {Richter}, \citenamefont {Hwang}, \citenamefont {Mannhart},\
  and\ \citenamefont {Moler}}]{Band1}%
  \BibitemOpen
  \bibfield  {author} {\bibinfo {author} {\bibfnamefont {B.}~\bibnamefont
  {Kalisky}}, \bibinfo {author} {\bibfnamefont {E.~M.}\ \bibnamefont
  {Spanton}}, \bibinfo {author} {\bibfnamefont {H.}~\bibnamefont {Noad}},
  \bibinfo {author} {\bibfnamefont {J.~R.}\ \bibnamefont {Kirtley}}, \bibinfo
  {author} {\bibfnamefont {K.~C.}\ \bibnamefont {Nowack}}, \bibinfo {author}
  {\bibfnamefont {C.}~\bibnamefont {Bell}}, \bibinfo {author} {\bibfnamefont
  {H.~K.}\ \bibnamefont {Sato}}, \bibinfo {author} {\bibfnamefont
  {M.}~\bibnamefont {Hosoda}}, \bibinfo {author} {\bibfnamefont
  {Y.}~\bibnamefont {Xie}}, \bibinfo {author} {\bibfnamefont {Y.}~\bibnamefont
  {Hikita}}, \bibinfo {author} {\bibfnamefont {C.}~\bibnamefont {Woltmann}},
  \bibinfo {author} {\bibfnamefont {G.}~\bibnamefont {Pfanzelt}}, \bibinfo
  {author} {\bibfnamefont {R.}~\bibnamefont {Jany}}, \bibinfo {author}
  {\bibfnamefont {C.}~\bibnamefont {Richter}}, \bibinfo {author} {\bibfnamefont
  {H.~Y.}\ \bibnamefont {Hwang}}, \bibinfo {author} {\bibfnamefont
  {J.}~\bibnamefont {Mannhart}}, \ and\ \bibinfo {author} {\bibfnamefont
  {K.~A.}\ \bibnamefont {Moler}},\ }\bibinfo {title} {{Locally enhanced
  conductivity due to the tetragonal domain structure in
  {L}a{A}l{O}$_3$/{S}r{T}i{O}$_3$ heterointerfaces}},\ \href {\doibase
  10.1038/nmat3753} {\bibfield  {journal} {\bibinfo  {journal} {Nat. Mater.}\
  }\textbf {\bibinfo {volume} {12}},\ \bibinfo {pages} {1091} (\bibinfo {year}
  {2013})}\BibitemShut {NoStop}%
\bibitem [{\citenamefont {Frenkel}\ \emph {et~al.}(2016)\citenamefont
  {Frenkel}, \citenamefont {Haham}, \citenamefont {Shperber}, \citenamefont
  {Bell}, \citenamefont {Xie}, \citenamefont {Chen}, \citenamefont {Hikita},
  \citenamefont {Hwang},\ and\ \citenamefont {Kalisky}}]{Band2}%
  \BibitemOpen
  \bibfield  {author} {\bibinfo {author} {\bibfnamefont {Y.}~\bibnamefont
  {Frenkel}}, \bibinfo {author} {\bibfnamefont {N.}~\bibnamefont {Haham}},
  \bibinfo {author} {\bibfnamefont {Y.}~\bibnamefont {Shperber}}, \bibinfo
  {author} {\bibfnamefont {C.}~\bibnamefont {Bell}}, \bibinfo {author}
  {\bibfnamefont {Y.}~\bibnamefont {Xie}}, \bibinfo {author} {\bibfnamefont
  {Z.}~\bibnamefont {Chen}}, \bibinfo {author} {\bibfnamefont {Y.}~\bibnamefont
  {Hikita}}, \bibinfo {author} {\bibfnamefont {H.~Y.}\ \bibnamefont {Hwang}}, \
  and\ \bibinfo {author} {\bibfnamefont {B.}~\bibnamefont {Kalisky}},\
  }\bibinfo {title} {{Anisotropic transport at the {L}a{A}l{O}$_3$
  /{S}r{T}i{O}$_3$ interface explained by microscopic imaging of channel-flow
  over {S}r{T}i{O}$_3$ domains}},\ \href {\doibase 10.1021/acsami.6b01655}
  {\bibfield  {journal} {\bibinfo  {journal} {ACS Appl. Mater. Interfaces}\
  }\textbf {\bibinfo {volume} {8}},\ \bibinfo {pages} {12514} (\bibinfo {year}
  {2016})}\BibitemShut {NoStop}%
\bibitem [{\citenamefont {Chang}\ \emph {et~al.}(2017)\citenamefont {Chang},
  \citenamefont {Eichler}, \citenamefont {Rhensius}, \citenamefont
  {Lorenzelli},\ and\ \citenamefont {Degen}}]{Chang}%
  \BibitemOpen
  \bibfield  {author} {\bibinfo {author} {\bibfnamefont {K.}~\bibnamefont
  {Chang}}, \bibinfo {author} {\bibfnamefont {A.}~\bibnamefont {Eichler}},
  \bibinfo {author} {\bibfnamefont {J.}~\bibnamefont {Rhensius}}, \bibinfo
  {author} {\bibfnamefont {L.}~\bibnamefont {Lorenzelli}}, \ and\ \bibinfo
  {author} {\bibfnamefont {C.~L.}\ \bibnamefont {Degen}},\ }\bibinfo {title}
  {{Nanoscale imaging of current density with a single-spin magnetometer}},\
  \href {\doibase 10.1021/acs.nanolett.6b05304} {\bibfield  {journal} {\bibinfo
   {journal} {Nano Lett.}\ }\textbf {\bibinfo {volume} {17}},\ \bibinfo {pages}
  {2367} (\bibinfo {year} {2017})}\BibitemShut {NoStop}%
\bibitem [{\citenamefont {Shadmi}\ \emph {et~al.}(2016)\citenamefont {Shadmi},
  \citenamefont {Kremen}, \citenamefont {Frenkel}, \citenamefont {Lapin},
  \citenamefont {Machado}, \citenamefont {Legoas}, \citenamefont {Bitton},
  \citenamefont {Rechav}, \citenamefont {Popovitz-Biro}, \citenamefont {Galva},
  \citenamefont {Jorio}, \citenamefont {Novotny}, \citenamefont {Kalisky},\
  and\ \citenamefont {Joselevich}}]{Shadmi}%
  \BibitemOpen
  \bibfield  {author} {\bibinfo {author} {\bibfnamefont {N.}~\bibnamefont
  {Shadmi}}, \bibinfo {author} {\bibfnamefont {A.}~\bibnamefont {Kremen}},
  \bibinfo {author} {\bibfnamefont {Y.}~\bibnamefont {Frenkel}}, \bibinfo
  {author} {\bibfnamefont {Z.~J.}\ \bibnamefont {Lapin}}, \bibinfo {author}
  {\bibfnamefont {L.~D.}\ \bibnamefont {Machado}}, \bibinfo {author}
  {\bibfnamefont {S.~B.}\ \bibnamefont {Legoas}}, \bibinfo {author}
  {\bibfnamefont {O.}~\bibnamefont {Bitton}}, \bibinfo {author} {\bibfnamefont
  {K.}~\bibnamefont {Rechav}}, \bibinfo {author} {\bibfnamefont
  {R.}~\bibnamefont {Popovitz-Biro}}, \bibinfo {author} {\bibfnamefont {D.~S.}\
  \bibnamefont {Galva}}, \bibinfo {author} {\bibfnamefont {A.}~\bibnamefont
  {Jorio}}, \bibinfo {author} {\bibfnamefont {L.}~\bibnamefont {Novotny}},
  \bibinfo {author} {\bibfnamefont {B.}~\bibnamefont {Kalisky}}, \ and\
  \bibinfo {author} {\bibfnamefont {E.}~\bibnamefont {Joselevich}},\ }\bibinfo
  {title} {{Defect-free carbon nanotube coils}},\ \href {\doibase
  10.1021/acs.nanolett.5b03417} {\bibfield  {journal} {\bibinfo  {journal}
  {Nano Lett.}\ }\textbf {\bibinfo {volume} {16}},\ \bibinfo {pages} {2152}
  (\bibinfo {year} {2016})}\BibitemShut {NoStop}%
\bibitem [{\citenamefont {Anahory}\ \emph {et~al.}(2014)\citenamefont
  {Anahory}, \citenamefont {Reiner}, \citenamefont {Embon}, \citenamefont
  {Halbertal}, \citenamefont {Yakovenko}, \citenamefont {Myasoedov},
  \citenamefont {Rappaport}, \citenamefont {Huber},\ and\ \citenamefont
  {Zeldov}}]{SOT5}%
  \BibitemOpen
  \bibfield  {author} {\bibinfo {author} {\bibfnamefont {Y.}~\bibnamefont
  {Anahory}}, \bibinfo {author} {\bibfnamefont {J.}~\bibnamefont {Reiner}},
  \bibinfo {author} {\bibfnamefont {L.}~\bibnamefont {Embon}}, \bibinfo
  {author} {\bibfnamefont {D.}~\bibnamefont {Halbertal}}, \bibinfo {author}
  {\bibfnamefont {A.}~\bibnamefont {Yakovenko}}, \bibinfo {author}
  {\bibfnamefont {Y.}~\bibnamefont {Myasoedov}}, \bibinfo {author}
  {\bibfnamefont {M.~L.}\ \bibnamefont {Rappaport}}, \bibinfo {author}
  {\bibfnamefont {M.~E.}\ \bibnamefont {Huber}}, \ and\ \bibinfo {author}
  {\bibfnamefont {E.}~\bibnamefont {Zeldov}},\ }\bibinfo {title}
  {{Three-junction SQUID-on-tip with tunable in-plane and out-of-plane magnetic
  field sensitivity}},\ \href {\doibase 10.1021/nl503022q} {\bibfield
  {journal} {\bibinfo  {journal} {Nano Lett.}\ }\textbf {\bibinfo {volume}
  {14}},\ \bibinfo {pages} {6481} (\bibinfo {year} {2014})}\BibitemShut
  {NoStop}%
\bibitem [{\citenamefont {Fleet}\ \emph {et~al.}(1999)\citenamefont {Fleet},
  \citenamefont {Chatraphorn},\ and\ \citenamefont {Wellstood}}]{Fleet1999}%
  \BibitemOpen
  \bibfield  {author} {\bibinfo {author} {\bibfnamefont {E.~F.}\ \bibnamefont
  {Fleet}}, \bibinfo {author} {\bibfnamefont {S.}~\bibnamefont {Chatraphorn}},
  \ and\ \bibinfo {author} {\bibfnamefont {F.~C.}\ \bibnamefont {Wellstood}},\
  }\bibinfo {title} {{HTS scanning SQUID microscopy of active circuits}},\
  \href {\doibase 10.1109/77.783928} {\bibfield  {journal} {\bibinfo  {journal}
  {IEEE Trans. Appl. Supercond.}\ }\textbf {\bibinfo {volume} {9}},\ \bibinfo
  {pages} {4103} (\bibinfo {year} {1999})}\BibitemShut {NoStop}%
\bibitem [{\citenamefont {Bending}(1999)}]{Bending1999}%
  \BibitemOpen
  \bibfield  {author} {\bibinfo {author} {\bibfnamefont {S.~J.}\ \bibnamefont
  {Bending}},\ }\bibinfo {title} {{Local magnetic probes of superconductors}},\
  \href {\doibase 10.1080/000187399243437} {\bibfield  {journal} {\bibinfo
  {journal} {Adv. Phys.}\ }\textbf {\bibinfo {volume} {48}},\ \bibinfo {pages}
  {449} (\bibinfo {year} {1999})}\BibitemShut {NoStop}%
\bibitem [{\citenamefont {Kirtley}(2010)}]{Kirtley2010}%
  \BibitemOpen
  \bibfield  {author} {\bibinfo {author} {\bibfnamefont {J.~R.}\ \bibnamefont
  {Kirtley}},\ }\bibinfo {title} {{Fundamental studies of superconductors using
  scanning magnetic imaging}},\ \href {\doibase 10.1088/0034-4885/73/12/126501}
  {\bibfield  {journal} {\bibinfo  {journal} {Reports Prog. Phys.}\ }\textbf
  {\bibinfo {volume} {73}},\ \bibinfo {pages} {126501} (\bibinfo {year}
  {2010})}\BibitemShut {NoStop}%
\bibitem [{\citenamefont {Grigorenko}\ \emph {et~al.}(2001)\citenamefont
  {Grigorenko}, \citenamefont {Bending}, \citenamefont {Tamegai}, \citenamefont
  {Ooi},\ and\ \citenamefont {Henini}}]{Grigorenko2001}%
  \BibitemOpen
  \bibfield  {author} {\bibinfo {author} {\bibfnamefont {A.}~\bibnamefont
  {Grigorenko}}, \bibinfo {author} {\bibfnamefont {S.}~\bibnamefont {Bending}},
  \bibinfo {author} {\bibfnamefont {T.}~\bibnamefont {Tamegai}}, \bibinfo
  {author} {\bibfnamefont {S.}~\bibnamefont {Ooi}}, \ and\ \bibinfo {author}
  {\bibfnamefont {M.}~\bibnamefont {Henini}},\ }\bibinfo {title} {{A
  one-dimensional chain state of vortex matter}},\ \href {\doibase
  10.1038/414728a} {\bibfield  {journal} {\bibinfo  {journal} {Nature}\
  }\textbf {\bibinfo {volume} {414}},\ \bibinfo {pages} {728} (\bibinfo {year}
  {2001})}\BibitemShut {NoStop}%
\bibitem [{\citenamefont {Kalisky}\ \emph {et~al.}(2009)\citenamefont
  {Kalisky}, \citenamefont {Kirtley}, \citenamefont {Nowadnick}, \citenamefont
  {Dinner}, \citenamefont {Zeldov}, \citenamefont {Ariando}, \citenamefont
  {Wenderich}, \citenamefont {Hilgenkamp}, \citenamefont {Feldmann},\ and\
  \citenamefont {Moler}}]{hp1}%
  \BibitemOpen
  \bibfield  {author} {\bibinfo {author} {\bibfnamefont {B.}~\bibnamefont
  {Kalisky}}, \bibinfo {author} {\bibfnamefont {J.~R.}\ \bibnamefont
  {Kirtley}}, \bibinfo {author} {\bibfnamefont {E.~A.}\ \bibnamefont
  {Nowadnick}}, \bibinfo {author} {\bibfnamefont {R.~B.}\ \bibnamefont
  {Dinner}}, \bibinfo {author} {\bibfnamefont {E.}~\bibnamefont {Zeldov}},
  \bibinfo {author} {\bibnamefont {Ariando}}, \bibinfo {author} {\bibfnamefont
  {S.}~\bibnamefont {Wenderich}}, \bibinfo {author} {\bibfnamefont
  {H.}~\bibnamefont {Hilgenkamp}}, \bibinfo {author} {\bibfnamefont {D.~M.}\
  \bibnamefont {Feldmann}}, \ and\ \bibinfo {author} {\bibfnamefont {K.~A.}\
  \bibnamefont {Moler}},\ }\bibinfo {title} {{Dynamics of single vortices in
  grain boundaries: I-V characteristics on the femtovolt scale}},\ \href
  {\doibase 10.1063/1.3137164} {\bibfield  {journal} {\bibinfo  {journal}
  {Appl. Phys. Lett.}\ }\textbf {\bibinfo {volume} {94}},\ \bibinfo {pages}
  {202504} (\bibinfo {year} {2009})}\BibitemShut {NoStop}%
\bibitem [{\citenamefont {Curran}\ \emph {et~al.}(2015)\citenamefont {Curran},
  \citenamefont {Kim}, \citenamefont {Satchell}, \citenamefont {Witt},
  \citenamefont {Burnell}, \citenamefont {Flokstra}, \citenamefont {Lee},
  \citenamefont {Cooper}, \citenamefont {Kinane}, \citenamefont {Langridge},
  \citenamefont {Isidori}, \citenamefont {Pugach}, \citenamefont {Eschrig},\
  and\ \citenamefont {Bending}}]{hp2}%
  \BibitemOpen
  \bibfield  {author} {\bibinfo {author} {\bibfnamefont {P.~J.}\ \bibnamefont
  {Curran}}, \bibinfo {author} {\bibfnamefont {J.}~\bibnamefont {Kim}},
  \bibinfo {author} {\bibfnamefont {N.}~\bibnamefont {Satchell}}, \bibinfo
  {author} {\bibfnamefont {J.~D.~S.}\ \bibnamefont {Witt}}, \bibinfo {author}
  {\bibfnamefont {G.}~\bibnamefont {Burnell}}, \bibinfo {author} {\bibfnamefont
  {M.~G.}\ \bibnamefont {Flokstra}}, \bibinfo {author} {\bibfnamefont {S.~L.}\
  \bibnamefont {Lee}}, \bibinfo {author} {\bibfnamefont {J.~F.~K.}\
  \bibnamefont {Cooper}}, \bibinfo {author} {\bibfnamefont {C.~J.}\
  \bibnamefont {Kinane}}, \bibinfo {author} {\bibfnamefont {S.}~\bibnamefont
  {Langridge}}, \bibinfo {author} {\bibfnamefont {A.}~\bibnamefont {Isidori}},
  \bibinfo {author} {\bibfnamefont {N.}~\bibnamefont {Pugach}}, \bibinfo
  {author} {\bibfnamefont {M.}~\bibnamefont {Eschrig}}, \ and\ \bibinfo
  {author} {\bibfnamefont {S.~J.}\ \bibnamefont {Bending}},\ }\bibinfo {title}
  {{Irreversible magnetization switching at the onset of superconductivity in a
  superconductor ferromagnet hybrid}},\ \href {\doibase 10.1063/1.4938467}
  {\bibfield  {journal} {\bibinfo  {journal} {Appl. Phys. Lett.}\ }\textbf
  {\bibinfo {volume} {107}},\ \bibinfo {pages} {1} (\bibinfo {year}
  {2015})}\BibitemShut {NoStop}%
\bibitem [{\citenamefont {Marchiori}\ \emph {et~al.}(2017)\citenamefont
  {Marchiori}, \citenamefont {Curra}, \citenamefont {Kim}, \citenamefont
  {Satchell}, \citenamefont {Burnell},\ and\ \citenamefont
  {Bending}}]{Marchiori2017}%
  \BibitemOpen
  \bibfield  {author} {\bibinfo {author} {\bibfnamefont {E.}~\bibnamefont
  {Marchiori}}, \bibinfo {author} {\bibfnamefont {P.~J.}\ \bibnamefont
  {Curra}}, \bibinfo {author} {\bibfnamefont {J.}~\bibnamefont {Kim}}, \bibinfo
  {author} {\bibfnamefont {N.}~\bibnamefont {Satchell}}, \bibinfo {author}
  {\bibfnamefont {G.}~\bibnamefont {Burnell}}, \ and\ \bibinfo {author}
  {\bibfnamefont {S.~J.}\ \bibnamefont {Bending}},\ }\bibinfo {title}
  {{Reconfigurable superconducting vortex pinning potential for magnetic disks
  in hybrid structures}},\ \href {\doibase 10.1038/srep45182} {\bibfield
  {journal} {\bibinfo  {journal} {Sci. Rep.}\ }\textbf {\bibinfo {volume}
  {7}},\ \bibinfo {pages} {45182} (\bibinfo {year} {2017})}\BibitemShut
  {NoStop}%
\bibitem [{\citenamefont {Beidenkopf}\ \emph {et~al.}(2005)\citenamefont
  {Beidenkopf}, \citenamefont {Avraham}, \citenamefont {Myasoedov},
  \citenamefont {Shtrikman}, \citenamefont {Zeldov}, \citenamefont
  {Rosenstein}, \citenamefont {Brandt},\ and\ \citenamefont {Tamegai}}]{HPA2}%
  \BibitemOpen
  \bibfield  {author} {\bibinfo {author} {\bibfnamefont {H.}~\bibnamefont
  {Beidenkopf}}, \bibinfo {author} {\bibfnamefont {N.}~\bibnamefont {Avraham}},
  \bibinfo {author} {\bibfnamefont {Y.}~\bibnamefont {Myasoedov}}, \bibinfo
  {author} {\bibfnamefont {H.}~\bibnamefont {Shtrikman}}, \bibinfo {author}
  {\bibfnamefont {E.}~\bibnamefont {Zeldov}}, \bibinfo {author} {\bibfnamefont
  {B.}~\bibnamefont {Rosenstein}}, \bibinfo {author} {\bibfnamefont {E.~H.}\
  \bibnamefont {Brandt}}, \ and\ \bibinfo {author} {\bibfnamefont
  {T.}~\bibnamefont {Tamegai}},\ }\bibinfo {title} {{Equilibrium first-order
  melting and second-order glass transitions of the vortex matter in
  Bi$_2$Sr$_2$CaCu$_2$O$_8$}},\ \href {\doibase 10.1103/PhysRevLett.95.257004}
  {\bibfield  {journal} {\bibinfo  {journal} {Phys. Rev. Lett.}\ }\textbf
  {\bibinfo {volume} {95}},\ \bibinfo {pages} {1} (\bibinfo {year}
  {2005})}\BibitemShut {NoStop}%
\bibitem [{\citenamefont {Paltiel}\ \emph {et~al.}(1999)\citenamefont
  {Paltiel}, \citenamefont {Zeldov}, \citenamefont {Myasoedov}, \citenamefont
  {Shtrikman}, \citenamefont {Bhattacharya}, \citenamefont {Higgins},
  \citenamefont {Xiao}, \citenamefont {Andrei}, \citenamefont {Gammel},\ and\
  \citenamefont {Bishop}}]{HPA3}%
  \BibitemOpen
  \bibfield  {author} {\bibinfo {author} {\bibfnamefont {Y.}~\bibnamefont
  {Paltiel}}, \bibinfo {author} {\bibfnamefont {E.}~\bibnamefont {Zeldov}},
  \bibinfo {author} {\bibfnamefont {Y.~N.}\ \bibnamefont {Myasoedov}}, \bibinfo
  {author} {\bibfnamefont {H.}~\bibnamefont {Shtrikman}}, \bibinfo {author}
  {\bibfnamefont {S.}~\bibnamefont {Bhattacharya}}, \bibinfo {author}
  {\bibfnamefont {M.~J.}\ \bibnamefont {Higgins}}, \bibinfo {author}
  {\bibfnamefont {Z.~L.}\ \bibnamefont {Xiao}}, \bibinfo {author}
  {\bibfnamefont {E.~Y.}\ \bibnamefont {Andrei}}, \bibinfo {author}
  {\bibfnamefont {P.~L.}\ \bibnamefont {Gammel}}, \ and\ \bibinfo {author}
  {\bibfnamefont {D.~J.}\ \bibnamefont {Bishop}},\ }\bibinfo {title} {{Dynamic
  instabilities and memory effects in vortex matter}},\ \href {\doibase
  10.1038/35000145} {\bibfield  {journal} {\bibinfo  {journal} {Nature}\
  }\textbf {\bibinfo {volume} {403}},\ \bibinfo {pages} {398} (\bibinfo {year}
  {1999})}\BibitemShut {NoStop}%
\bibitem [{\citenamefont {Baruch-El}\ \emph {et~al.}(2016)\citenamefont
  {Baruch-El}, \citenamefont {Baziljevich}, \citenamefont {Shapiro},
  \citenamefont {Johansen}, \citenamefont {Shaulov},\ and\ \citenamefont
  {Yeshurun}}]{MO1}%
  \BibitemOpen
  \bibfield  {author} {\bibinfo {author} {\bibfnamefont {E.}~\bibnamefont
  {Baruch-El}}, \bibinfo {author} {\bibfnamefont {M.}~\bibnamefont
  {Baziljevich}}, \bibinfo {author} {\bibfnamefont {B.~Y.}\ \bibnamefont
  {Shapiro}}, \bibinfo {author} {\bibfnamefont {T.~H.}\ \bibnamefont
  {Johansen}}, \bibinfo {author} {\bibfnamefont {A.}~\bibnamefont {Shaulov}}, \
  and\ \bibinfo {author} {\bibfnamefont {Y.}~\bibnamefont {Yeshurun}},\
  }\bibinfo {title} {{Dendritic flux instabilities in YBa$_2$Cu$_3$O$_{7-x}$
  films : Effects of temperature and magnetic field ramp rate}},\ \href
  {\doibase 10.1103/PhysRevB.94.054509} {\bibfield  {journal} {\bibinfo
  {journal} {Phys. Rev. B}\ }\textbf {\bibinfo {volume} {94}},\ \bibinfo
  {pages} {054509} (\bibinfo {year} {2016})}\BibitemShut {NoStop}%
\bibitem [{\citenamefont {Albrecht}\ \emph {et~al.}(2016)\citenamefont
  {Albrecht}, \citenamefont {Br{\"{u}}ck}, \citenamefont {Stahl},\ and\
  \citenamefont {Ruo{\ss}}}]{MO2}%
  \BibitemOpen
  \bibfield  {author} {\bibinfo {author} {\bibfnamefont {J.}~\bibnamefont
  {Albrecht}}, \bibinfo {author} {\bibfnamefont {S.}~\bibnamefont
  {Br{\"{u}}ck}}, \bibinfo {author} {\bibfnamefont {C.}~\bibnamefont {Stahl}},
  \ and\ \bibinfo {author} {\bibfnamefont {S.}~\bibnamefont {Ruo{\ss}}},\
  }\bibinfo {title} {{Quantitative magneto-optical analysis of the role of
  finite temperatures on the critical state in YBCO thin films}},\ \href
  {\doibase 10.1088/0953-2048/29/11/114002} {\bibfield  {journal} {\bibinfo
  {journal} {Supercond. Sci. Technol.}\ }\textbf {\bibinfo {volume} {29}},\
  \bibinfo {pages} {114002} (\bibinfo {year} {2016})}\BibitemShut {NoStop}%
\bibitem [{\citenamefont {Yuan}\ \emph {et~al.}(2016)\citenamefont {Yuan},
  \citenamefont {Xu}, \citenamefont {Ma}, \citenamefont {Sun},\ and\
  \citenamefont {Tamegai}}]{MO3}%
  \BibitemOpen
  \bibfield  {author} {\bibinfo {author} {\bibfnamefont {P.}~\bibnamefont
  {Yuan}}, \bibinfo {author} {\bibfnamefont {Z.}~\bibnamefont {Xu}}, \bibinfo
  {author} {\bibfnamefont {Y.}~\bibnamefont {Ma}}, \bibinfo {author}
  {\bibfnamefont {Y.}~\bibnamefont {Sun}}, \ and\ \bibinfo {author}
  {\bibfnamefont {T.}~\bibnamefont {Tamegai}},\ }\bibinfo {title}
  {{Angular-dependent vortex pinning mechanism and magneto-optical
  characterizations of FeSe$_{0.5}$Te$_{0.5}$ thin films grown on CaF$_2$
  substrates}},\ \href {\doibase 10.1088/0953-2048/29/3/035013} {\bibfield
  {journal} {\bibinfo  {journal} {Supercond. Sci. Technol.}\ }\textbf {\bibinfo
  {volume} {29}},\ \bibinfo {pages} {035013} (\bibinfo {year}
  {2016})}\BibitemShut {NoStop}%
\bibitem [{\citenamefont {Vlasko-Vlasov}\ \emph {et~al.}(2015)\citenamefont
  {Vlasko-Vlasov}, \citenamefont {Glatz}, \citenamefont {Koshelev},
  \citenamefont {Welp},\ and\ \citenamefont {Kwok}}]{MO4}%
  \BibitemOpen
  \bibfield  {author} {\bibinfo {author} {\bibfnamefont {V.~K.}\ \bibnamefont
  {Vlasko-Vlasov}}, \bibinfo {author} {\bibfnamefont {A.}~\bibnamefont
  {Glatz}}, \bibinfo {author} {\bibfnamefont {A.~E.}\ \bibnamefont {Koshelev}},
  \bibinfo {author} {\bibfnamefont {U.}~\bibnamefont {Welp}}, \ and\ \bibinfo
  {author} {\bibfnamefont {W.~K.}\ \bibnamefont {Kwok}},\ }\bibinfo {title}
  {{Anisotropic superconductors in tilted magnetic fields}},\ \href {\doibase
  10.1103/PhysRevB.91.224505} {\bibfield  {journal} {\bibinfo  {journal} {Phys.
  Rev. B}\ }\textbf {\bibinfo {volume} {91}},\ \bibinfo {pages} {224505}
  (\bibinfo {year} {2015})}\BibitemShut {NoStop}%
\bibitem [{\citenamefont {Baziljevich}\ \emph {et~al.}(2014)\citenamefont
  {Baziljevich}, \citenamefont {Baruch-El}, \citenamefont {Johansen},\ and\
  \citenamefont {Yeshurun}}]{MO5}%
  \BibitemOpen
  \bibfield  {author} {\bibinfo {author} {\bibfnamefont {M.}~\bibnamefont
  {Baziljevich}}, \bibinfo {author} {\bibfnamefont {E.}~\bibnamefont
  {Baruch-El}}, \bibinfo {author} {\bibfnamefont {T.~H.}\ \bibnamefont
  {Johansen}}, \ and\ \bibinfo {author} {\bibfnamefont {Y.}~\bibnamefont
  {Yeshurun}},\ }\bibinfo {title} {{Dendritic instability in
  YBa$_2$Cu$_3$O$_{7-\delta}$ films triggered by transient magnetic fields}},\
  \href {\doibase 10.1063/1.4887374} {\bibfield  {journal} {\bibinfo  {journal}
  {Appl. Phys. Lett.}\ }\textbf {\bibinfo {volume} {105}},\ \bibinfo {pages}
  {012602} (\bibinfo {year} {2014})}\BibitemShut {NoStop}%
\bibitem [{\citenamefont {Fang}\ \emph {et~al.}(2013)\citenamefont {Fang},
  \citenamefont {Jia}, \citenamefont {Mishra}, \citenamefont {Chaparro},
  \citenamefont {Vlasko-Vlasov}, \citenamefont {Koshelev}, \citenamefont
  {Welp}, \citenamefont {Crabtree}, \citenamefont {Zhu}, \citenamefont
  {Zhigadlo}, \citenamefont {Katrych}, \citenamefont {Karpinski},\ and\
  \citenamefont {Kwok}}]{MO6}%
  \BibitemOpen
  \bibfield  {author} {\bibinfo {author} {\bibfnamefont {L.}~\bibnamefont
  {Fang}}, \bibinfo {author} {\bibfnamefont {Y.}~\bibnamefont {Jia}}, \bibinfo
  {author} {\bibfnamefont {V.}~\bibnamefont {Mishra}}, \bibinfo {author}
  {\bibfnamefont {C.}~\bibnamefont {Chaparro}}, \bibinfo {author}
  {\bibfnamefont {V.~K.}\ \bibnamefont {Vlasko-Vlasov}}, \bibinfo {author}
  {\bibfnamefont {A.~E.}\ \bibnamefont {Koshelev}}, \bibinfo {author}
  {\bibfnamefont {U.}~\bibnamefont {Welp}}, \bibinfo {author} {\bibfnamefont
  {G.~W.}\ \bibnamefont {Crabtree}}, \bibinfo {author} {\bibfnamefont
  {S.}~\bibnamefont {Zhu}}, \bibinfo {author} {\bibfnamefont {N.~D.}\
  \bibnamefont {Zhigadlo}}, \bibinfo {author} {\bibfnamefont {S.}~\bibnamefont
  {Katrych}}, \bibinfo {author} {\bibfnamefont {J.}~\bibnamefont {Karpinski}},
  \ and\ \bibinfo {author} {\bibfnamefont {W.~K.}\ \bibnamefont {Kwok}},\
  }\bibinfo {title} {{Huge critical current density and tailored
  superconducting anisotropy in SmFeAsO$_{0.8}$F$_{0.15}$ by low-density
  columnar-defect incorporation}},\ \href {\doibase 10.1038/ncomms3655}
  {\bibfield  {journal} {\bibinfo  {journal} {Nat. Commun.}\ }\textbf {\bibinfo
  {volume} {4}},\ \bibinfo {pages} {2655} (\bibinfo {year} {2013})}\BibitemShut
  {NoStop}%
\bibitem [{\citenamefont {Prozorov}\ \emph {et~al.}(2008)\citenamefont
  {Prozorov}, \citenamefont {Fidler}, \citenamefont {Hoberg},\ and\
  \citenamefont {Canfield}}]{MO7}%
  \BibitemOpen
  \bibfield  {author} {\bibinfo {author} {\bibfnamefont {R.}~\bibnamefont
  {Prozorov}}, \bibinfo {author} {\bibfnamefont {A.~F.}\ \bibnamefont
  {Fidler}}, \bibinfo {author} {\bibfnamefont {J.~R.}\ \bibnamefont {Hoberg}},
  \ and\ \bibinfo {author} {\bibfnamefont {P.~C.}\ \bibnamefont {Canfield}},\
  }\bibinfo {title} {{Suprafroth in type-I superconductors}},\ \href {\doibase
  10.1038/nphys888} {\bibfield  {journal} {\bibinfo  {journal} {Nat. Phys.}\
  }\textbf {\bibinfo {volume} {4}},\ \bibinfo {pages} {327} (\bibinfo {year}
  {2008})}\BibitemShut {NoStop}%
\bibitem [{\citenamefont {Kalisky}\ \emph {et~al.}(2007)\citenamefont
  {Kalisky}, \citenamefont {Myasoedov}, \citenamefont {Shaulov}, \citenamefont
  {Tamegai}, \citenamefont {Zeldov},\ and\ \citenamefont {Yeshurun}}]{MO8}%
  \BibitemOpen
  \bibfield  {author} {\bibinfo {author} {\bibfnamefont {B.}~\bibnamefont
  {Kalisky}}, \bibinfo {author} {\bibfnamefont {Y.}~\bibnamefont {Myasoedov}},
  \bibinfo {author} {\bibfnamefont {A.}~\bibnamefont {Shaulov}}, \bibinfo
  {author} {\bibfnamefont {T.}~\bibnamefont {Tamegai}}, \bibinfo {author}
  {\bibfnamefont {E.}~\bibnamefont {Zeldov}}, \ and\ \bibinfo {author}
  {\bibfnamefont {Y.}~\bibnamefont {Yeshurun}},\ }\bibinfo {title} {{Dynamic
  order-to-metastable-disorder vortex matter transition in
  Bi$_2$Sr$_2$CaCu$_2$O$_{8+\delta}$}},\ \href {\doibase
  10.1103/PhysRevLett.98.107001} {\bibfield  {journal} {\bibinfo  {journal}
  {Phys. Rev. Lett.}\ }\textbf {\bibinfo {volume} {98}},\ \bibinfo {pages}
  {107001} (\bibinfo {year} {2007})}\BibitemShut {NoStop}%
\bibitem [{\citenamefont {Soibel}\ \emph {et~al.}(2000)\citenamefont {Soibel},
  \citenamefont {Zeldov}, \citenamefont {Rappaport}, \citenamefont {Myasoedov},
  \citenamefont {Tamegai}, \citenamefont {Ooi}, \citenamefont {Konczykowski},\
  and\ \citenamefont {Geshkenbein}}]{MO9}%
  \BibitemOpen
  \bibfield  {author} {\bibinfo {author} {\bibfnamefont {A.}~\bibnamefont
  {Soibel}}, \bibinfo {author} {\bibfnamefont {E.}~\bibnamefont {Zeldov}},
  \bibinfo {author} {\bibfnamefont {M.}~\bibnamefont {Rappaport}}, \bibinfo
  {author} {\bibfnamefont {Y.}~\bibnamefont {Myasoedov}}, \bibinfo {author}
  {\bibfnamefont {T.}~\bibnamefont {Tamegai}}, \bibinfo {author} {\bibfnamefont
  {S.}~\bibnamefont {Ooi}}, \bibinfo {author} {\bibfnamefont {M.}~\bibnamefont
  {Konczykowski}}, \ and\ \bibinfo {author} {\bibfnamefont {V.~B.}\
  \bibnamefont {Geshkenbein}},\ }\bibinfo {title} {{Imaging the vortex-lattice
  melting process in the presence of disorder}},\ \href {\doibase
  10.1038/35018532} {\bibfield  {journal} {\bibinfo  {journal} {Nature}\
  }\textbf {\bibinfo {volume} {406}},\ \bibinfo {pages} {282} (\bibinfo {year}
  {2000})}\BibitemShut {NoStop}%
\bibitem [{\citenamefont {Veauvy}\ \emph {et~al.}(2002)\citenamefont {Veauvy},
  \citenamefont {Hasselbach},\ and\ \citenamefont {Mailly}}]{comp2}%
  \BibitemOpen
  \bibfield  {author} {\bibinfo {author} {\bibfnamefont {C.}~\bibnamefont
  {Veauvy}}, \bibinfo {author} {\bibfnamefont {K.}~\bibnamefont {Hasselbach}},
  \ and\ \bibinfo {author} {\bibfnamefont {D.}~\bibnamefont {Mailly}},\
  }\bibinfo {title} {{Scanning $\mu$-superconduction quantum interference
  device force microscope}},\ \href {\doibase 10.1063/1.1515384} {\bibfield
  {journal} {\bibinfo  {journal} {Rev. Sci. Instrum.}\ }\textbf {\bibinfo
  {volume} {73}},\ \bibinfo {pages} {3825} (\bibinfo {year}
  {2002})}\BibitemShut {NoStop}%
\bibitem [{\citenamefont {Koshnick}\ \emph {et~al.}(2008)\citenamefont
  {Koshnick}, \citenamefont {Huber}, \citenamefont {Bert}, \citenamefont
  {Hicks}, \citenamefont {Large}, \citenamefont {Edwards},\ and\ \citenamefont
  {Moler}}]{comp1}%
  \BibitemOpen
  \bibfield  {author} {\bibinfo {author} {\bibfnamefont {N.~C.}\ \bibnamefont
  {Koshnick}}, \bibinfo {author} {\bibfnamefont {M.~E.}\ \bibnamefont {Huber}},
  \bibinfo {author} {\bibfnamefont {J.~A.}\ \bibnamefont {Bert}}, \bibinfo
  {author} {\bibfnamefont {C.~W.}\ \bibnamefont {Hicks}}, \bibinfo {author}
  {\bibfnamefont {J.}~\bibnamefont {Large}}, \bibinfo {author} {\bibfnamefont
  {H.}~\bibnamefont {Edwards}}, \ and\ \bibinfo {author} {\bibfnamefont
  {K.~A.}\ \bibnamefont {Moler}},\ }\bibinfo {title} {{A terraced scanning
  supper conducting quantum interference device susceptometer with submicron
  pickup loops}},\ \href {\doibase 10.1063/1.3046098} {\bibfield  {journal}
  {\bibinfo  {journal} {Appl. Phys. Lett.}\ }\textbf {\bibinfo {volume} {93}},\
  \bibinfo {pages} {243101} (\bibinfo {year} {2008})}\BibitemShut {NoStop}%
\bibitem [{\citenamefont {Huber}\ \emph {et~al.}(2008)\citenamefont {Huber},
  \citenamefont {Koshnick}, \citenamefont {Bluhm}, \citenamefont {Archuleta},
  \citenamefont {Azua}, \citenamefont {Bj{\"{o}}rnsson}, \citenamefont
  {Gardner}, \citenamefont {Halloran}, \citenamefont {Lucero},\ and\
  \citenamefont {Moler}}]{SCANNING2}%
  \BibitemOpen
  \bibfield  {author} {\bibinfo {author} {\bibfnamefont {M.~E.}\ \bibnamefont
  {Huber}}, \bibinfo {author} {\bibfnamefont {N.~C.}\ \bibnamefont {Koshnick}},
  \bibinfo {author} {\bibfnamefont {H.}~\bibnamefont {Bluhm}}, \bibinfo
  {author} {\bibfnamefont {L.~J.}\ \bibnamefont {Archuleta}}, \bibinfo {author}
  {\bibfnamefont {T.}~\bibnamefont {Azua}}, \bibinfo {author} {\bibfnamefont
  {P.~G.}\ \bibnamefont {Bj{\"{o}}rnsson}}, \bibinfo {author} {\bibfnamefont
  {B.~W.}\ \bibnamefont {Gardner}}, \bibinfo {author} {\bibfnamefont {S.~T.}\
  \bibnamefont {Halloran}}, \bibinfo {author} {\bibfnamefont {E.~A.}\
  \bibnamefont {Lucero}}, \ and\ \bibinfo {author} {\bibfnamefont {K.~A.}\
  \bibnamefont {Moler}},\ }\bibinfo {title} {{Gradiometric micro-SQUID
  susceptometer for scanning measurements of mesoscopic samples}},\ \href
  {\doibase 10.1063/1.2932341} {\bibfield  {journal} {\bibinfo  {journal} {Rev.
  Sci. Instrum.}\ }\textbf {\bibinfo {volume} {79}},\ \bibinfo {pages} {053704}
  (\bibinfo {year} {2008})}\BibitemShut {NoStop}%
\bibitem [{\citenamefont {Talanov}\ \emph {et~al.}(2014)\citenamefont
  {Talanov}, \citenamefont {{Lettsome Jr}}, \citenamefont {Borzenets},
  \citenamefont {Gagliolo}, \citenamefont {Cawthorne},\ and\ \citenamefont
  {Orozco}}]{SCANNING4}%
  \BibitemOpen
  \bibfield  {author} {\bibinfo {author} {\bibfnamefont {V.~V.}\ \bibnamefont
  {Talanov}}, \bibinfo {author} {\bibfnamefont {N.~M.}\ \bibnamefont {{Lettsome
  Jr}}}, \bibinfo {author} {\bibfnamefont {V.}~\bibnamefont {Borzenets}},
  \bibinfo {author} {\bibfnamefont {N.}~\bibnamefont {Gagliolo}}, \bibinfo
  {author} {\bibfnamefont {A.~B.}\ \bibnamefont {Cawthorne}}, \ and\ \bibinfo
  {author} {\bibfnamefont {A.}~\bibnamefont {Orozco}},\ }\bibinfo {title} {{A
  scanning SQUID microscope with 200 MHz bandwidth}},\ \href {\doibase
  10.1088/0953-2048/27/4/044032} {\bibfield  {journal} {\bibinfo  {journal}
  {Supercond. Sci. Technol.}\ }\textbf {\bibinfo {volume} {27}},\ \bibinfo
  {pages} {044032} (\bibinfo {year} {2014})}\BibitemShut {NoStop}%
\bibitem [{\citenamefont {Walbrecker}\ \emph {et~al.}(2014)\citenamefont
  {Walbrecker}, \citenamefont {Kalisky}, \citenamefont {Grombacher},
  \citenamefont {Kirtley}, \citenamefont {Moler},\ and\ \citenamefont
  {Knight}}]{SCANNING6}%
  \BibitemOpen
  \bibfield  {author} {\bibinfo {author} {\bibfnamefont {J.~O.}\ \bibnamefont
  {Walbrecker}}, \bibinfo {author} {\bibfnamefont {B.}~\bibnamefont {Kalisky}},
  \bibinfo {author} {\bibfnamefont {D.}~\bibnamefont {Grombacher}}, \bibinfo
  {author} {\bibfnamefont {J.}~\bibnamefont {Kirtley}}, \bibinfo {author}
  {\bibfnamefont {K.~A.}\ \bibnamefont {Moler}}, \ and\ \bibinfo {author}
  {\bibfnamefont {R.}~\bibnamefont {Knight}},\ }\bibinfo {title} {{Direct
  measurement of internal magnetic fields in natural sands using scanning SQUID
  microscopy}},\ \href {\doibase 10.1016/j.jmr.2014.01.012} {\bibfield
  {journal} {\bibinfo  {journal} {J. Magn. Reson.}\ }\textbf {\bibinfo {volume}
  {242}},\ \bibinfo {pages} {10} (\bibinfo {year} {2014})}\BibitemShut
  {NoStop}%
\bibitem [{\citenamefont {Hykel}\ \emph {et~al.}(2014)\citenamefont {Hykel},
  \citenamefont {Wang}, \citenamefont {Castellazzi}, \citenamefont {Crozes},
  \citenamefont {Shaw}, \citenamefont {Schuster},\ and\ \citenamefont
  {Hasselbach}}]{Hykel2014}%
  \BibitemOpen
  \bibfield  {author} {\bibinfo {author} {\bibfnamefont {D.~J.}\ \bibnamefont
  {Hykel}}, \bibinfo {author} {\bibfnamefont {Z.~S.}\ \bibnamefont {Wang}},
  \bibinfo {author} {\bibfnamefont {P.}~\bibnamefont {Castellazzi}}, \bibinfo
  {author} {\bibfnamefont {T.}~\bibnamefont {Crozes}}, \bibinfo {author}
  {\bibfnamefont {G.}~\bibnamefont {Shaw}}, \bibinfo {author} {\bibfnamefont
  {K.}~\bibnamefont {Schuster}}, \ and\ \bibinfo {author} {\bibfnamefont
  {K.}~\bibnamefont {Hasselbach}},\ }\bibinfo {title} {{MicroSQUID force
  microscopy in a dilution refrigerator}},\ \href {\doibase
  10.1007/s10909-014-1174-9} {\bibfield  {journal} {\bibinfo  {journal} {J. Low
  Temp. Phys.}\ }\textbf {\bibinfo {volume} {175}},\ \bibinfo {pages} {861}
  (\bibinfo {year} {2014})}\BibitemShut {NoStop}%
\bibitem [{\citenamefont {Wang}\ \emph {et~al.}(2015)\citenamefont {Wang},
  \citenamefont {Li}, \citenamefont {L{\"{u}}}, \citenamefont {Paudel},
  \citenamefont {Leusink}, \citenamefont {Hoek}, \citenamefont {Poccia},
  \citenamefont {Vailionis}, \citenamefont {Venkatesan}, \citenamefont {Coey},
  \citenamefont {Tsymbal}, \citenamefont {Ariando},\ and\ \citenamefont
  {Hilgenkamp}}]{SCANNING9}%
  \BibitemOpen
  \bibfield  {author} {\bibinfo {author} {\bibfnamefont {X.~R.}\ \bibnamefont
  {Wang}}, \bibinfo {author} {\bibfnamefont {C.~J.}\ \bibnamefont {Li}},
  \bibinfo {author} {\bibfnamefont {W.~M.}\ \bibnamefont {L{\"{u}}}}, \bibinfo
  {author} {\bibfnamefont {T.~R.}\ \bibnamefont {Paudel}}, \bibinfo {author}
  {\bibfnamefont {D.~P.}\ \bibnamefont {Leusink}}, \bibinfo {author}
  {\bibfnamefont {M.}~\bibnamefont {Hoek}}, \bibinfo {author} {\bibfnamefont
  {N.}~\bibnamefont {Poccia}}, \bibinfo {author} {\bibfnamefont
  {A.}~\bibnamefont {Vailionis}}, \bibinfo {author} {\bibfnamefont
  {T.}~\bibnamefont {Venkatesan}}, \bibinfo {author} {\bibfnamefont {J.~M.~D.}\
  \bibnamefont {Coey}}, \bibinfo {author} {\bibfnamefont {E.~Y.}\ \bibnamefont
  {Tsymbal}}, \bibinfo {author} {\bibnamefont {Ariando}}, \ and\ \bibinfo
  {author} {\bibfnamefont {H.}~\bibnamefont {Hilgenkamp}},\ }\bibinfo {title}
  {{Imaging and control of ferromagnetism in {L}a{M}n{O}$_3$/{S}r{T}i{O}$_3$
  heterostructures}},\ \href {\doibase 10.1126/science.aaa5198} {\bibfield
  {journal} {\bibinfo  {journal} {Science}\ }\textbf {\bibinfo {volume}
  {349}},\ \bibinfo {pages} {716} (\bibinfo {year} {2015})}\BibitemShut
  {NoStop}%
\bibitem [{\citenamefont {Kremen}\ \emph {et~al.}(2016)\citenamefont {Kremen},
  \citenamefont {Wissberg}, \citenamefont {Haham}, \citenamefont {Persky},
  \citenamefont {Frenkel},\ and\ \citenamefont {Kalisky}}]{Kremen2016}%
  \BibitemOpen
  \bibfield  {author} {\bibinfo {author} {\bibfnamefont {A.}~\bibnamefont
  {Kremen}}, \bibinfo {author} {\bibfnamefont {S.}~\bibnamefont {Wissberg}},
  \bibinfo {author} {\bibfnamefont {N.}~\bibnamefont {Haham}}, \bibinfo
  {author} {\bibfnamefont {E.}~\bibnamefont {Persky}}, \bibinfo {author}
  {\bibfnamefont {Y.}~\bibnamefont {Frenkel}}, \ and\ \bibinfo {author}
  {\bibfnamefont {B.}~\bibnamefont {Kalisky}},\ }\bibinfo {title} {{Mechanical
  control of individual superconducting vortices}},\ \href {\doibase
  10.1021/acs.nanolett.5b04444} {\bibfield  {journal} {\bibinfo  {journal}
  {Nano Lett.}\ }\textbf {\bibinfo {volume} {16}},\ \bibinfo {pages} {1626}
  (\bibinfo {year} {2016})}\BibitemShut {NoStop}%
\bibitem [{\citenamefont {Maletinsky}\ \emph {et~al.}(2012)\citenamefont
  {Maletinsky}, \citenamefont {Hong}, \citenamefont {Grinolds}, \citenamefont
  {Hausmann}, \citenamefont {Lukin}, \citenamefont {Walsworth}, \citenamefont
  {Loncar},\ and\ \citenamefont {Yacoby}}]{Maletinsky}%
  \BibitemOpen
  \bibfield  {author} {\bibinfo {author} {\bibfnamefont {P.}~\bibnamefont
  {Maletinsky}}, \bibinfo {author} {\bibfnamefont {S.}~\bibnamefont {Hong}},
  \bibinfo {author} {\bibfnamefont {M.~S.}\ \bibnamefont {Grinolds}}, \bibinfo
  {author} {\bibfnamefont {B.}~\bibnamefont {Hausmann}}, \bibinfo {author}
  {\bibfnamefont {M.~D.}\ \bibnamefont {Lukin}}, \bibinfo {author}
  {\bibfnamefont {R.~L.}\ \bibnamefont {Walsworth}}, \bibinfo {author}
  {\bibfnamefont {M.}~\bibnamefont {Loncar}}, \ and\ \bibinfo {author}
  {\bibfnamefont {A.}~\bibnamefont {Yacoby}},\ }\bibinfo {title} {{A robust
  scanning diamond sensor for nanoscale imaging with single nitrogen-vacancy
  centres}},\ \href {\doibase 10.1038/nnano.2012.50} {\bibfield  {journal}
  {\bibinfo  {journal} {Nat. Nanotechnol.}\ }\textbf {\bibinfo {volume} {7}},\
  \bibinfo {pages} {320} (\bibinfo {year} {2012})}\BibitemShut {NoStop}%
\bibitem [{\citenamefont {Pelliccione}\ \emph {et~al.}(2016)\citenamefont
  {Pelliccione}, \citenamefont {Jenkins}, \citenamefont {Ovartchaiyapong},
  \citenamefont {Reetz}, \citenamefont {Emmanuelidu}, \citenamefont {Ni},\ and\
  \citenamefont {{Bleszynski-Jayich}}}]{NV1}%
  \BibitemOpen
  \bibfield  {author} {\bibinfo {author} {\bibfnamefont {M.}~\bibnamefont
  {Pelliccione}}, \bibinfo {author} {\bibfnamefont {A.}~\bibnamefont
  {Jenkins}}, \bibinfo {author} {\bibfnamefont {P.}~\bibnamefont
  {Ovartchaiyapong}}, \bibinfo {author} {\bibfnamefont {C.}~\bibnamefont
  {Reetz}}, \bibinfo {author} {\bibfnamefont {E.}~\bibnamefont {Emmanuelidu}},
  \bibinfo {author} {\bibfnamefont {N.}~\bibnamefont {Ni}}, \ and\ \bibinfo
  {author} {\bibfnamefont {A.~C.}\ \bibnamefont {{Bleszynski-Jayich}}},\
  }\bibinfo {title} {{Scanned probe imaging of nanoscale magnetism at cryogenic
  temperatures with a single-spin quantum sensor}},\ \href {\doibase
  10.1038/nnano.2016.68} {\bibfield  {journal} {\bibinfo  {journal} {Nat.
  Nanotechnol.}\ }\textbf {\bibinfo {volume} {11}},\ \bibinfo {pages} {700}
  (\bibinfo {year} {2016})}\BibitemShut {NoStop}%
\bibitem [{\citenamefont {Thiel}\ \emph {et~al.}(2016)\citenamefont {Thiel},
  \citenamefont {Rohner}, \citenamefont {Ganzhorn}, \citenamefont {Appel},
  \citenamefont {Neu}, \citenamefont {M{\"{u}}ller}, \citenamefont {Kleiner},
  \citenamefont {Koelle},\ and\ \citenamefont {Maletinsky}}]{NV2}%
  \BibitemOpen
  \bibfield  {author} {\bibinfo {author} {\bibfnamefont {L.}~\bibnamefont
  {Thiel}}, \bibinfo {author} {\bibfnamefont {D.}~\bibnamefont {Rohner}},
  \bibinfo {author} {\bibfnamefont {M.}~\bibnamefont {Ganzhorn}}, \bibinfo
  {author} {\bibfnamefont {P.}~\bibnamefont {Appel}}, \bibinfo {author}
  {\bibfnamefont {E.}~\bibnamefont {Neu}}, \bibinfo {author} {\bibfnamefont
  {B.}~\bibnamefont {M{\"{u}}ller}}, \bibinfo {author} {\bibfnamefont
  {R.}~\bibnamefont {Kleiner}}, \bibinfo {author} {\bibfnamefont
  {D.}~\bibnamefont {Koelle}}, \ and\ \bibinfo {author} {\bibfnamefont
  {P.}~\bibnamefont {Maletinsky}},\ }\bibinfo {title} {{Quantitative nanoscale
  vortex-imaging using a cryogenic quantum magnetometer}},\ \href {\doibase
  10.1038/nnano.2016.63} {\bibfield  {journal} {\bibinfo  {journal} {Nat.
  Nanotechnol.}\ }\textbf {\bibinfo {volume} {11}},\ \bibinfo {pages} {677}
  (\bibinfo {year} {2016})}\BibitemShut {NoStop}%
\bibitem [{\citenamefont {Dovzhenko}\ \emph {et~al.}()\citenamefont
  {Dovzhenko}, \citenamefont {Casola}, \citenamefont {Schlotter}, \citenamefont
  {Zhou}, \citenamefont {Walsworth}, \citenamefont {Beach},\ and\ \citenamefont
  {Yacoby}}]{Dovzhenko}%
  \BibitemOpen
  \bibfield  {author} {\bibinfo {author} {\bibfnamefont {Y.}~\bibnamefont
  {Dovzhenko}}, \bibinfo {author} {\bibfnamefont {F.}~\bibnamefont {Casola}},
  \bibinfo {author} {\bibfnamefont {S.}~\bibnamefont {Schlotter}}, \bibinfo
  {author} {\bibfnamefont {T.~X.}\ \bibnamefont {Zhou}}, \bibinfo {author}
  {\bibfnamefont {R.~L.}\ \bibnamefont {Walsworth}}, \bibinfo {author}
  {\bibfnamefont {G.~S.~D.}\ \bibnamefont {Beach}}, \ and\ \bibinfo {author}
  {\bibfnamefont {A.}~\bibnamefont {Yacoby}},\ }\bibinfo {title} {{Imaging the
  spin texture of a skyrmion under ambient conditions using an atomic-sized
  sensor}},\ \href@noop {} {\bibinfo  {journal} {ArXiv e-prints,
  https://arxiv.org/abs/1611.00673}\ }\BibitemShut {NoStop}%
\bibitem [{\citenamefont {Granata}\ and\ \citenamefont
  {Vettoliere}(2016)}]{Granata2015}%
  \BibitemOpen
\bibfield  {journal} {  }\bibfield  {author} {\bibinfo {author} {\bibfnamefont
  {C.}~\bibnamefont {Granata}}\ and\ \bibinfo {author} {\bibfnamefont
  {A.}~\bibnamefont {Vettoliere}},\ }\bibinfo {title} {{Nano superconducting
  quantum interference device: a powerful tool for nanoscale investigations}},\
  \href {\doibase 10.1016/j.physrep.2015.12.001} {\bibfield  {journal}
  {\bibinfo  {journal} {Phys. Rep.}\ }\textbf {\bibinfo {volume} {614}},\
  \bibinfo {pages} {1} (\bibinfo {year} {2016})}\BibitemShut {NoStop}%
\bibitem [{\citenamefont {Vasyukov}\ \emph {et~al.}(2013)\citenamefont
  {Vasyukov}, \citenamefont {Anahory}, \citenamefont {Embon}, \citenamefont
  {Halbertal}, \citenamefont {Cuppens}, \citenamefont {Neeman}, \citenamefont
  {Finkler}, \citenamefont {Segev}, \citenamefont {Myasoedov}, \citenamefont
  {Rappaport}, \citenamefont {Huber},\ and\ \citenamefont {Zeldov}}]{SOT1}%
  \BibitemOpen
  \bibfield  {author} {\bibinfo {author} {\bibfnamefont {D.}~\bibnamefont
  {Vasyukov}}, \bibinfo {author} {\bibfnamefont {Y.}~\bibnamefont {Anahory}},
  \bibinfo {author} {\bibfnamefont {L.}~\bibnamefont {Embon}}, \bibinfo
  {author} {\bibfnamefont {D.}~\bibnamefont {Halbertal}}, \bibinfo {author}
  {\bibfnamefont {J.}~\bibnamefont {Cuppens}}, \bibinfo {author} {\bibfnamefont
  {L.}~\bibnamefont {Neeman}}, \bibinfo {author} {\bibfnamefont
  {A.}~\bibnamefont {Finkler}}, \bibinfo {author} {\bibfnamefont
  {Y.}~\bibnamefont {Segev}}, \bibinfo {author} {\bibfnamefont
  {Y.}~\bibnamefont {Myasoedov}}, \bibinfo {author} {\bibfnamefont {M.~L.}\
  \bibnamefont {Rappaport}}, \bibinfo {author} {\bibfnamefont {M.~E.}\
  \bibnamefont {Huber}}, \ and\ \bibinfo {author} {\bibfnamefont
  {E.}~\bibnamefont {Zeldov}},\ }\bibinfo {title} {{A scanning superconducting
  quantum interference device with single electron spin sensitivity}},\ \href
  {\doibase 10.1038/nnano.2013.169} {\bibfield  {journal} {\bibinfo  {journal}
  {Nat. Nanotechnol.}\ }\textbf {\bibinfo {volume} {8}},\ \bibinfo {pages}
  {639} (\bibinfo {year} {2013})}\BibitemShut {NoStop}%
\bibitem [{\citenamefont {Nagel}\ \emph {et~al.}(2013)\citenamefont {Nagel},
  \citenamefont {Buchter}, \citenamefont {Xue}, \citenamefont {Kieler},
  \citenamefont {Weimann}, \citenamefont {Kohlmann}, \citenamefont {Zorin},
  \citenamefont {R{\"{u}}ffer}, \citenamefont {Russo-Averchi}, \citenamefont
  {Huber}, \citenamefont {Berberich}, \citenamefont {{Fontcuberta i Morral}},
  \citenamefont {Grundler}, \citenamefont {Kleiner}, \citenamefont {Koelle},
  \citenamefont {Poggio},\ and\ \citenamefont {Kemmler}}]{SCANNING3}%
  \BibitemOpen
  \bibfield  {author} {\bibinfo {author} {\bibfnamefont {J.}~\bibnamefont
  {Nagel}}, \bibinfo {author} {\bibfnamefont {A.}~\bibnamefont {Buchter}},
  \bibinfo {author} {\bibfnamefont {F.}~\bibnamefont {Xue}}, \bibinfo {author}
  {\bibfnamefont {O.~F.}\ \bibnamefont {Kieler}}, \bibinfo {author}
  {\bibfnamefont {T.}~\bibnamefont {Weimann}}, \bibinfo {author} {\bibfnamefont
  {J.}~\bibnamefont {Kohlmann}}, \bibinfo {author} {\bibfnamefont {A.~B.}\
  \bibnamefont {Zorin}}, \bibinfo {author} {\bibfnamefont {D.}~\bibnamefont
  {R{\"{u}}ffer}}, \bibinfo {author} {\bibfnamefont {E.}~\bibnamefont
  {Russo-Averchi}}, \bibinfo {author} {\bibfnamefont {R.}~\bibnamefont
  {Huber}}, \bibinfo {author} {\bibfnamefont {P.}~\bibnamefont {Berberich}},
  \bibinfo {author} {\bibfnamefont {A.}~\bibnamefont {{Fontcuberta i Morral}}},
  \bibinfo {author} {\bibfnamefont {D.}~\bibnamefont {Grundler}}, \bibinfo
  {author} {\bibfnamefont {R.}~\bibnamefont {Kleiner}}, \bibinfo {author}
  {\bibfnamefont {D.}~\bibnamefont {Koelle}}, \bibinfo {author} {\bibfnamefont
  {M.}~\bibnamefont {Poggio}}, \ and\ \bibinfo {author} {\bibfnamefont
  {M.}~\bibnamefont {Kemmler}},\ }\bibinfo {title} {{Nanoscale multifunctional
  sensor formed by a Ni nanotube and a scanning Nb nanoSQUID}},\ \href
  {\doibase 10.1103/PhysRevB.88.064425} {\bibfield  {journal} {\bibinfo
  {journal} {Phys. Rev. B}\ }\textbf {\bibinfo {volume} {88}},\ \bibinfo
  {pages} {064425} (\bibinfo {year} {2013})}\BibitemShut {NoStop}%
\bibitem [{\citenamefont {Finkler}\ \emph {et~al.}(2010)\citenamefont
  {Finkler}, \citenamefont {Segev}, \citenamefont {Myasoedov}, \citenamefont
  {Rappaport}, \citenamefont {Ne'Eman}, \citenamefont {Vasyukov}, \citenamefont
  {Zeldov}, \citenamefont {Huber}, \citenamefont {Martin},\ and\ \citenamefont
  {Yacoby}}]{SOT2}%
  \BibitemOpen
  \bibfield  {author} {\bibinfo {author} {\bibfnamefont {A.}~\bibnamefont
  {Finkler}}, \bibinfo {author} {\bibfnamefont {Y.}~\bibnamefont {Segev}},
  \bibinfo {author} {\bibfnamefont {Y.}~\bibnamefont {Myasoedov}}, \bibinfo
  {author} {\bibfnamefont {M.~L.}\ \bibnamefont {Rappaport}}, \bibinfo {author}
  {\bibfnamefont {L.}~\bibnamefont {Ne'Eman}}, \bibinfo {author} {\bibfnamefont
  {D.}~\bibnamefont {Vasyukov}}, \bibinfo {author} {\bibfnamefont
  {E.}~\bibnamefont {Zeldov}}, \bibinfo {author} {\bibfnamefont {M.~E.}\
  \bibnamefont {Huber}}, \bibinfo {author} {\bibfnamefont {J.}~\bibnamefont
  {Martin}}, \ and\ \bibinfo {author} {\bibfnamefont {A.}~\bibnamefont
  {Yacoby}},\ }\bibinfo {title} {{Self-aligned nanoscale SQUID on a tip}},\
  \href {\doibase 10.1021/nl100009r} {\bibfield  {journal} {\bibinfo  {journal}
  {Nano Lett.}\ }\textbf {\bibinfo {volume} {10}},\ \bibinfo {pages} {1046}
  (\bibinfo {year} {2010})}\BibitemShut {NoStop}%
\bibitem [{\citenamefont {Embon}\ \emph {et~al.}(2015)\citenamefont {Embon},
  \citenamefont {Anahory}, \citenamefont {Suhov}, \citenamefont {Halbertal},
  \citenamefont {Cuppens}, \citenamefont {Yakovenko}, \citenamefont {Uri},
  \citenamefont {Myasoedov}, \citenamefont {Rappaport}, \citenamefont {Huber},
  \citenamefont {Gurevich},\ and\ \citenamefont {Zeldov}}]{SCANNING8}%
  \BibitemOpen
  \bibfield  {author} {\bibinfo {author} {\bibfnamefont {L.}~\bibnamefont
  {Embon}}, \bibinfo {author} {\bibfnamefont {Y.}~\bibnamefont {Anahory}},
  \bibinfo {author} {\bibfnamefont {A.}~\bibnamefont {Suhov}}, \bibinfo
  {author} {\bibfnamefont {D.}~\bibnamefont {Halbertal}}, \bibinfo {author}
  {\bibfnamefont {J.}~\bibnamefont {Cuppens}}, \bibinfo {author} {\bibfnamefont
  {A.}~\bibnamefont {Yakovenko}}, \bibinfo {author} {\bibfnamefont
  {A.}~\bibnamefont {Uri}}, \bibinfo {author} {\bibfnamefont {Y.}~\bibnamefont
  {Myasoedov}}, \bibinfo {author} {\bibfnamefont {M.~L.}\ \bibnamefont
  {Rappaport}}, \bibinfo {author} {\bibfnamefont {M.~E.}\ \bibnamefont
  {Huber}}, \bibinfo {author} {\bibfnamefont {A.}~\bibnamefont {Gurevich}}, \
  and\ \bibinfo {author} {\bibfnamefont {E.}~\bibnamefont {Zeldov}},\ }\bibinfo
  {title} {{Probing dynamics and pinning of single vortices in superconductors
  at nanometer scales}},\ \href {\doibase 10.1038/srep07598} {\bibfield
  {journal} {\bibinfo  {journal} {Sci. Rep.}\ }\textbf {\bibinfo {volume}
  {5}},\ \bibinfo {pages} {7598} (\bibinfo {year} {2015})}\BibitemShut
  {NoStop}%
\bibitem [{\citenamefont {Lachman}\ \emph {et~al.}(2015)\citenamefont
  {Lachman}, \citenamefont {Young}, \citenamefont {Richardella}, \citenamefont
  {Cuppens}, \citenamefont {Naren}, \citenamefont {Anahory}, \citenamefont
  {Meltzer}, \citenamefont {Kandala}, \citenamefont {Kempinger}, \citenamefont
  {Myasoedov}, \citenamefont {Huber}, \citenamefont {Samarth},\ and\
  \citenamefont {Zeldov}}]{Lachman2015}%
  \BibitemOpen
  \bibfield  {author} {\bibinfo {author} {\bibfnamefont {E.~O.}\ \bibnamefont
  {Lachman}}, \bibinfo {author} {\bibfnamefont {A.~F.}\ \bibnamefont {Young}},
  \bibinfo {author} {\bibfnamefont {A.}~\bibnamefont {Richardella}}, \bibinfo
  {author} {\bibfnamefont {J.}~\bibnamefont {Cuppens}}, \bibinfo {author}
  {\bibfnamefont {H.}~\bibnamefont {Naren}}, \bibinfo {author} {\bibfnamefont
  {Y.}~\bibnamefont {Anahory}}, \bibinfo {author} {\bibfnamefont {A.~Y.}\
  \bibnamefont {Meltzer}}, \bibinfo {author} {\bibfnamefont {A.}~\bibnamefont
  {Kandala}}, \bibinfo {author} {\bibfnamefont {S.}~\bibnamefont {Kempinger}},
  \bibinfo {author} {\bibfnamefont {Y.}~\bibnamefont {Myasoedov}}, \bibinfo
  {author} {\bibfnamefont {M.~E.}\ \bibnamefont {Huber}}, \bibinfo {author}
  {\bibfnamefont {N.}~\bibnamefont {Samarth}}, \ and\ \bibinfo {author}
  {\bibfnamefont {E.}~\bibnamefont {Zeldov}},\ }\bibinfo {title}
  {{Visualization of superparamagnetic dynamics in magnetic topological
  insulators.}}\ \href {\doibase 10.1126/sciadv.1500740} {\bibfield  {journal}
  {\bibinfo  {journal} {Sci. Adv.}\ }\textbf {\bibinfo {volume} {1}},\ \bibinfo
  {pages} {e1500740} (\bibinfo {year} {2015})}\BibitemShut {NoStop}%
\bibitem [{\citenamefont {Shibata}\ \emph {et~al.}(2015)\citenamefont
  {Shibata}, \citenamefont {Nomura}, \citenamefont {Kashiwaya}, \citenamefont
  {Kashiwaya}, \citenamefont {Ishiguro},\ and\ \citenamefont
  {Takayanagi}}]{SCANNING1}%
  \BibitemOpen
  \bibfield  {author} {\bibinfo {author} {\bibfnamefont {Y.}~\bibnamefont
  {Shibata}}, \bibinfo {author} {\bibfnamefont {S.}~\bibnamefont {Nomura}},
  \bibinfo {author} {\bibfnamefont {H.}~\bibnamefont {Kashiwaya}}, \bibinfo
  {author} {\bibfnamefont {S.}~\bibnamefont {Kashiwaya}}, \bibinfo {author}
  {\bibfnamefont {R.}~\bibnamefont {Ishiguro}}, \ and\ \bibinfo {author}
  {\bibfnamefont {H.}~\bibnamefont {Takayanagi}},\ }\bibinfo {title} {{Imaging
  of current density distributions with a Nb weak-link scanning nano-SQUID
  microscope}},\ \href {\doibase 10.1038/srep15097} {\bibfield  {journal}
  {\bibinfo  {journal} {Sci. Rep.}\ }\textbf {\bibinfo {volume} {5}},\ \bibinfo
  {pages} {15097} (\bibinfo {year} {2015})}\BibitemShut {NoStop}%
\bibitem [{\citenamefont {Anahory}\ \emph {et~al.}(2016)\citenamefont
  {Anahory}, \citenamefont {Embon}, \citenamefont {Li}, \citenamefont
  {Banerjee}, \citenamefont {Meltzer}, \citenamefont {Naren}, \citenamefont
  {Yakovenko}, \citenamefont {Cuppens}, \citenamefont {Myasoedov},
  \citenamefont {Rappaport}, \citenamefont {Huber}, \citenamefont {Michaeli},
  \citenamefont {Venkatesan}, \citenamefont {Ariando},\ and\ \citenamefont
  {Zeldov}}]{SCANNING5}%
  \BibitemOpen
  \bibfield  {author} {\bibinfo {author} {\bibfnamefont {Y.}~\bibnamefont
  {Anahory}}, \bibinfo {author} {\bibfnamefont {L.}~\bibnamefont {Embon}},
  \bibinfo {author} {\bibfnamefont {C.~J.}\ \bibnamefont {Li}}, \bibinfo
  {author} {\bibfnamefont {S.}~\bibnamefont {Banerjee}}, \bibinfo {author}
  {\bibfnamefont {A.~Y.}\ \bibnamefont {Meltzer}}, \bibinfo {author}
  {\bibfnamefont {H.~R.}\ \bibnamefont {Naren}}, \bibinfo {author}
  {\bibfnamefont {A.}~\bibnamefont {Yakovenko}}, \bibinfo {author}
  {\bibfnamefont {J.}~\bibnamefont {Cuppens}}, \bibinfo {author} {\bibfnamefont
  {Y.}~\bibnamefont {Myasoedov}}, \bibinfo {author} {\bibfnamefont {M.~L.}\
  \bibnamefont {Rappaport}}, \bibinfo {author} {\bibfnamefont {M.~E.}\
  \bibnamefont {Huber}}, \bibinfo {author} {\bibfnamefont {K.}~\bibnamefont
  {Michaeli}}, \bibinfo {author} {\bibfnamefont {T.}~\bibnamefont
  {Venkatesan}}, \bibinfo {author} {\bibnamefont {Ariando}}, \ and\ \bibinfo
  {author} {\bibfnamefont {E.}~\bibnamefont {Zeldov}},\ }\bibinfo {title}
  {{Emergent nanoscale superparamagnetism at oxide interfaces}},\ \href
  {\doibase 10.1038/ncomms12566} {\bibfield  {journal} {\bibinfo  {journal}
  {Nat. Commun.}\ }\textbf {\bibinfo {volume} {7}},\ \bibinfo {pages} {12566}
  (\bibinfo {year} {2016})}\BibitemShut {NoStop}%
\bibitem [{\citenamefont {Uri}\ \emph {et~al.}(2016)\citenamefont {Uri},
  \citenamefont {Meltzer}, \citenamefont {Anahory}, \citenamefont {Embon},
  \citenamefont {Lachman}, \citenamefont {Halbertal}, \citenamefont {Hr},
  \citenamefont {Myasoedov}, \citenamefont {Huber}, \citenamefont {Young},\
  and\ \citenamefont {Zeldov}}]{Aviram}%
  \BibitemOpen
  \bibfield  {author} {\bibinfo {author} {\bibfnamefont {A.}~\bibnamefont
  {Uri}}, \bibinfo {author} {\bibfnamefont {A.~Y.}\ \bibnamefont {Meltzer}},
  \bibinfo {author} {\bibfnamefont {Y.}~\bibnamefont {Anahory}}, \bibinfo
  {author} {\bibfnamefont {L.}~\bibnamefont {Embon}}, \bibinfo {author}
  {\bibfnamefont {E.~O.}\ \bibnamefont {Lachman}}, \bibinfo {author}
  {\bibfnamefont {D.}~\bibnamefont {Halbertal}}, \bibinfo {author}
  {\bibfnamefont {N.}~\bibnamefont {Hr}}, \bibinfo {author} {\bibfnamefont
  {Y.}~\bibnamefont {Myasoedov}}, \bibinfo {author} {\bibfnamefont {M.~E.}\
  \bibnamefont {Huber}}, \bibinfo {author} {\bibfnamefont {A.~F.}\ \bibnamefont
  {Young}}, \ and\ \bibinfo {author} {\bibfnamefont {E.}~\bibnamefont
  {Zeldov}},\ }\bibinfo {title} {{Electrically tunable multiterminal
  SQUID-on-tip}},\ \href {\doibase 10.1021/acs.nanolett.6b02841} {\bibfield
  {journal} {\bibinfo  {journal} {Nano Lett.}\ }\textbf {\bibinfo {volume}
  {16}},\ \bibinfo {pages} {6910} (\bibinfo {year} {2016})}\BibitemShut
  {NoStop}%
\bibitem [{\citenamefont {Wildermuth}\ \emph {et~al.}(2005)\citenamefont
  {Wildermuth}, \citenamefont {Hofferberth}, \citenamefont {Lesanovsky},
  \citenamefont {Haller}, \citenamefont {Andersson}, \citenamefont {Groth},
  \citenamefont {Bar-Joseph}, \citenamefont {Kruger},\ and\ \citenamefont
  {Schmiedmayer}}]{AC1}%
  \BibitemOpen
  \bibfield  {author} {\bibinfo {author} {\bibfnamefont {S.}~\bibnamefont
  {Wildermuth}}, \bibinfo {author} {\bibfnamefont {S.}~\bibnamefont
  {Hofferberth}}, \bibinfo {author} {\bibfnamefont {I.}~\bibnamefont
  {Lesanovsky}}, \bibinfo {author} {\bibfnamefont {E.}~\bibnamefont {Haller}},
  \bibinfo {author} {\bibfnamefont {L.~M.}\ \bibnamefont {Andersson}}, \bibinfo
  {author} {\bibfnamefont {S.}~\bibnamefont {Groth}}, \bibinfo {author}
  {\bibfnamefont {I.}~\bibnamefont {Bar-Joseph}}, \bibinfo {author}
  {\bibfnamefont {P.}~\bibnamefont {Kruger}}, \ and\ \bibinfo {author}
  {\bibfnamefont {J.}~\bibnamefont {Schmiedmayer}},\ }\bibinfo {title}
  {{Bose–Einstein condensates: Microscopic magnetic-field imaging}},\ \href
  {\doibase 10.1038/435440a} {\bibfield  {journal} {\bibinfo  {journal}
  {Nature}\ }\textbf {\bibinfo {volume} {435}},\ \bibinfo {pages} {440}
  (\bibinfo {year} {2005})}\BibitemShut {NoStop}%
\bibitem [{\citenamefont {Aigner}\ \emph {et~al.}(2008)\citenamefont {Aigner},
  \citenamefont {Pietra}, \citenamefont {Japha}, \citenamefont {David},
  \citenamefont {Salem}, \citenamefont {Folman},\ and\ \citenamefont
  {Schmiedmayer}}]{Aigner}%
  \BibitemOpen
  \bibfield  {author} {\bibinfo {author} {\bibfnamefont {S.}~\bibnamefont
  {Aigner}}, \bibinfo {author} {\bibfnamefont {L.~D.}\ \bibnamefont {Pietra}},
  \bibinfo {author} {\bibfnamefont {Y.}~\bibnamefont {Japha}}, \bibinfo
  {author} {\bibfnamefont {T.}~\bibnamefont {David}}, \bibinfo {author}
  {\bibfnamefont {R.}~\bibnamefont {Salem}}, \bibinfo {author} {\bibfnamefont
  {R.}~\bibnamefont {Folman}}, \ and\ \bibinfo {author} {\bibfnamefont
  {J.}~\bibnamefont {Schmiedmayer}},\ }\bibinfo {title} {{Long-range order in
  electronic transport through disordered metal films}},\ \href {\doibase
  10.1126/science.1152458} {\bibfield  {journal} {\bibinfo  {journal}
  {Science}\ }\textbf {\bibinfo {volume} {319}},\ \bibinfo {pages} {1226}
  (\bibinfo {year} {2008})}\BibitemShut {NoStop}%
\bibitem [{\citenamefont {Yang}\ \emph {et~al.}(2017)\citenamefont {Yang},
  \citenamefont {Koll{\'{a}}r}, \citenamefont {Taylor}, \citenamefont
  {Turner},\ and\ \citenamefont {Lev}}]{Yang}%
  \BibitemOpen
  \bibfield  {author} {\bibinfo {author} {\bibfnamefont {F.}~\bibnamefont
  {Yang}}, \bibinfo {author} {\bibfnamefont {A.~J.}\ \bibnamefont
  {Koll{\'{a}}r}}, \bibinfo {author} {\bibfnamefont {S.~F.}\ \bibnamefont
  {Taylor}}, \bibinfo {author} {\bibfnamefont {R.~W.}\ \bibnamefont {Turner}},
  \ and\ \bibinfo {author} {\bibfnamefont {B.~L.}\ \bibnamefont {Lev}},\
  }\bibinfo {title} {{Scanning Quantum Cryogenic Atom Microscope}},\ \href
  {\doibase 10.1103/PhysRevApplied.7.034026} {\bibfield  {journal} {\bibinfo
  {journal} {Phys. Rev. Appl.}\ }\textbf {\bibinfo {volume} {7}},\ \bibinfo
  {pages} {034026} (\bibinfo {year} {2017})}\BibitemShut {NoStop}%
\bibitem [{\citenamefont {Roth}\ \emph {et~al.}(1989)\citenamefont {Roth},
  \citenamefont {Sepulveda},\ and\ \citenamefont {Wikswo}}]{exactFT}%
  \BibitemOpen
  \bibfield  {author} {\bibinfo {author} {\bibfnamefont {B.~J.}\ \bibnamefont
  {Roth}}, \bibinfo {author} {\bibfnamefont {N.~G.}\ \bibnamefont {Sepulveda}},
  \ and\ \bibinfo {author} {\bibfnamefont {J.~P.}\ \bibnamefont {Wikswo}},\
  }\bibinfo {title} {{Using a magnetometer to image a two-dimensional current
  distribution}},\ \href {\doibase 10.1063/1.342549} {\bibfield  {journal}
  {\bibinfo  {journal} {J. Appl. Phys.}\ }\textbf {\bibinfo {volume} {65}},\
  \bibinfo {pages} {361} (\bibinfo {year} {1989})}\BibitemShut {NoStop}%
\bibitem [{\citenamefont {Wijngaarden}\ \emph {et~al.}(1998)\citenamefont
  {Wijngaarden}, \citenamefont {Heeck}, \citenamefont {Spoelder}, \citenamefont
  {Surdeanu},\ and\ \citenamefont {Griessen}}]{CG-FFT}%
  \BibitemOpen
  \bibfield  {author} {\bibinfo {author} {\bibfnamefont {R.~J.}\ \bibnamefont
  {Wijngaarden}}, \bibinfo {author} {\bibfnamefont {K.}~\bibnamefont {Heeck}},
  \bibinfo {author} {\bibfnamefont {H.~J.~W.}\ \bibnamefont {Spoelder}},
  \bibinfo {author} {\bibfnamefont {R.}~\bibnamefont {Surdeanu}}, \ and\
  \bibinfo {author} {\bibfnamefont {R.}~\bibnamefont {Griessen}},\ }\bibinfo
  {title} {{Fast determination of 2D current patterns in flat conductors from
  measurement of their magnetic field}},\ \href {\doibase
  10.1016/S0921-4534(97)01799-1} {\bibfield  {journal} {\bibinfo  {journal}
  {Phys. C Supercond.}\ }\textbf {\bibinfo {volume} {295}},\ \bibinfo {pages}
  {177} (\bibinfo {year} {1998})}\BibitemShut {NoStop}%
\bibitem [{\citenamefont {Wijngaarden}\ \emph {et~al.}(1996)\citenamefont
  {Wijngaarden}, \citenamefont {Spoelder}, \citenamefont {Surdeanu},\ and\
  \citenamefont {Griessen}}]{CG}%
  \BibitemOpen
  \bibfield  {author} {\bibinfo {author} {\bibfnamefont {R.~J.}\ \bibnamefont
  {Wijngaarden}}, \bibinfo {author} {\bibfnamefont {H.~J.~W.}\ \bibnamefont
  {Spoelder}}, \bibinfo {author} {\bibfnamefont {R.}~\bibnamefont {Surdeanu}},
  \ and\ \bibinfo {author} {\bibfnamefont {R.}~\bibnamefont {Griessen}},\
  }\bibinfo {title} {{Determination of two-dimensional current patterns in flat
  superconductors from magneto-optical measurements: an efficient inversion
  scheme}},\ \href {\doibase 10.1103/PhysRevB.54.6742} {\bibfield  {journal}
  {\bibinfo  {journal} {Phys. Rev. B}\ }\textbf {\bibinfo {volume} {54}},\
  \bibinfo {pages} {6742} (\bibinfo {year} {1996})}\BibitemShut {NoStop}%
\bibitem [{\citenamefont {Feldmann}(2004)}]{Feldman}%
  \BibitemOpen
  \bibfield  {author} {\bibinfo {author} {\bibfnamefont {D.~M.}\ \bibnamefont
  {Feldmann}},\ }\bibinfo {title} {{Resolution of two-dimensional currents in
  superconductors from a two-dimensional magnetic field measurement by the
  method of regularization}},\ \href {\doibase 10.1103/PhysRevB.69.144515}
  {\bibfield  {journal} {\bibinfo  {journal} {Phys. Rev. B}\ }\textbf {\bibinfo
  {volume} {69}},\ \bibinfo {pages} {144515} (\bibinfo {year}
  {2004})}\BibitemShut {NoStop}%
\bibitem [{\citenamefont {Lavrentiev}(1967)}]{MMLavrentiev1967}%
  \BibitemOpen
  \bibfield  {author} {\bibinfo {author} {\bibfnamefont {M.~M.}\ \bibnamefont
  {Lavrentiev}},\ }\href {\doibase 10.1007/978-3-642-88210-4} {\emph {\bibinfo
  {title} {{Some Improperly Posed Problems of Mathematical Physics}}}},\
  \bibinfo {series} {Springer Tracts in Natural Philosophy}, Vol.~\bibinfo
  {volume} {11}\ (\bibinfo  {publisher} {Springer},\ \bibinfo {address}
  {Berlin, Heidelberg},\ \bibinfo {year} {1967})\BibitemShut {NoStop}%
\bibitem [{\citenamefont {Tikhonov}\ and\ \citenamefont
  {Arsenin}(1977)}]{tikhonov1977solutions}%
  \BibitemOpen
  \bibfield  {author} {\bibinfo {author} {\bibfnamefont {A.~N.}\ \bibnamefont
  {Tikhonov}}\ and\ \bibinfo {author} {\bibfnamefont {V.~Y.}\ \bibnamefont
  {Arsenin}},\ }\href@noop {} {\emph {\bibinfo {title} {{Solutions of Ill-Posed
  Problems}}}}\ (\bibinfo  {publisher} {John Wiley \& Sons Inc},\ \bibinfo
  {year} {1977})\BibitemShut {NoStop}%
\bibitem [{\citenamefont {Wahba}(1990)}]{wahba1990spline}%
  \BibitemOpen
  \bibfield  {author} {\bibinfo {author} {\bibfnamefont {G.}~\bibnamefont
  {Wahba}},\ }\href {\doibase 10.1137/1.9781611970128} {\emph {\bibinfo {title}
  {{Spline Models for Observational Data}}}},\ Vol.~\bibinfo {volume} {59}\
  (\bibinfo  {publisher} {SIAM},\ \bibinfo {address} {Philadelphia},\ \bibinfo
  {year} {1990})\BibitemShut {NoStop}%
\bibitem [{\citenamefont {Stein}(1981)}]{Stein1981}%
  \BibitemOpen
  \bibfield  {author} {\bibinfo {author} {\bibfnamefont {C.~M.}\ \bibnamefont
  {Stein}},\ }\bibinfo {title} {{Estimation of the mean of a multivariate
  normal distribution}},\ \href {\doibase 10.1214/aos/1176345632} {\bibfield
  {journal} {\bibinfo  {journal} {Ann. Stat.}\ }\textbf {\bibinfo {volume}
  {9}},\ \bibinfo {pages} {1135} (\bibinfo {year} {1981})}\BibitemShut
  {NoStop}%
\bibitem [{\citenamefont {Ramani}\ \emph {et~al.}(2008)\citenamefont {Ramani},
  \citenamefont {Blu},\ and\ \citenamefont {Unser}}]{Ramani2008}%
  \BibitemOpen
  \bibfield  {author} {\bibinfo {author} {\bibfnamefont {S.}~\bibnamefont
  {Ramani}}, \bibinfo {author} {\bibfnamefont {T.}~\bibnamefont {Blu}}, \ and\
  \bibinfo {author} {\bibfnamefont {M.}~\bibnamefont {Unser}},\ }\bibinfo
  {title} {{Monte-Carlo SURE: a black-box optimization of regularization
  parameters for general denoising algorithms}},\ \href {\doibase
  10.1109/TIP.2008.2001404} {\bibfield  {journal} {\bibinfo  {journal} {IEEE
  Trans. Image Process.}\ }\textbf {\bibinfo {volume} {17}},\ \bibinfo {pages}
  {1540} (\bibinfo {year} {2008})}\BibitemShut {NoStop}%
\bibitem [{\citenamefont {Ramani}\ \emph {et~al.}(2012)\citenamefont {Ramani},
  \citenamefont {Liu}, \citenamefont {Rosen}, \citenamefont {Nielsen},\ and\
  \citenamefont {Fessler}}]{ProjectedSure}%
  \BibitemOpen
  \bibfield  {author} {\bibinfo {author} {\bibfnamefont {S.}~\bibnamefont
  {Ramani}}, \bibinfo {author} {\bibfnamefont {Z.}~\bibnamefont {Liu}},
  \bibinfo {author} {\bibfnamefont {J.}~\bibnamefont {Rosen}}, \bibinfo
  {author} {\bibfnamefont {J.-F.}\ \bibnamefont {Nielsen}}, \ and\ \bibinfo
  {author} {\bibfnamefont {J.~A.}\ \bibnamefont {Fessler}},\ }\bibinfo {title}
  {{Regularization parameter selection for nonlinear iterative image
  restoration and MRI reconstruction using GCV and SURE-based methods}},\ \href
  {\doibase 10.1109/TIP.2012.2195015} {\bibfield  {journal} {\bibinfo
  {journal} {IEEE Trans. Image Process.}\ }\textbf {\bibinfo {volume} {21}},\
  \bibinfo {pages} {3659} (\bibinfo {year} {2012})}\BibitemShut {NoStop}%
\bibitem [{Note1()}]{Note1}%
  \BibitemOpen
  \bibinfo {note} {A MATLAB-based implementation of the described algorithms
  and the numerical examples used in this paper can be found at:
  {https://www.weizmann.ac.il/condmat/superc/software/}.}\BibitemShut {Stop}%
\bibitem [{\citenamefont {Levin}\ and\ \citenamefont {Meltzer}()}]{PP2016}%
  \BibitemOpen
  \bibfield  {author} {\bibinfo {author} {\bibfnamefont {E.}~\bibnamefont
  {Levin}}\ and\ \bibinfo {author} {\bibfnamefont {A.~Y.}\ \bibnamefont
  {Meltzer}},\ }\bibinfo {title} {{Estimation of the regularization parameter
  in linear discrete ill-posed problems using the Picard parameter}},\
  \href@noop {} {\bibinfo  {journal} {ArXiv e-prints,
  https://arxiv.org/abs/1607.00938}\ }\BibitemShut {NoStop}%
\bibitem [{\citenamefont {Press}\ \emph {et~al.}(1992)\citenamefont {Press},
  \citenamefont {Teukolsky}, \citenamefont {Vetterling},\ and\ \citenamefont
  {Flannery}}]{pressnumerical}%
  \BibitemOpen
\bibfield  {journal} {  }\bibfield  {author} {\bibinfo {author} {\bibfnamefont
  {W.~H.}\ \bibnamefont {Press}}, \bibinfo {author} {\bibfnamefont {S.~A.}\
  \bibnamefont {Teukolsky}}, \bibinfo {author} {\bibfnamefont {W.~T.}\
  \bibnamefont {Vetterling}}, \ and\ \bibinfo {author} {\bibfnamefont {B.~P.}\
  \bibnamefont {Flannery}},\ }\href@noop {} {\bibinfo {title} {{Numerical
  Recipes in FORTRAN: The Art of Scientific Computing}},\ } (\bibinfo {year}
  {1992})\BibitemShut {NoStop}%
\bibitem [{Note2()}]{Note2}%
  \BibitemOpen
  \bibinfo {note} {Note, that the formula for GCV given in Equation 15 in \cite
  {Feldman} contains a typographical error.}\BibitemShut {Stop}%
\bibitem [{\citenamefont {O'Leary}(2001)}]{NearOpt2006}%
  \BibitemOpen
  \bibfield  {author} {\bibinfo {author} {\bibfnamefont {D.~P.}\ \bibnamefont
  {O'Leary}},\ }\bibinfo {title} {{Near-optimal parameters for Tikhonov and
  other regularization methods}},\ \href {\doibase 10.1137/S1064827599354147}
  {\bibfield  {journal} {\bibinfo  {journal} {SIAM J. Sci. Comput.}\ }\textbf
  {\bibinfo {volume} {23}},\ \bibinfo {pages} {1161} (\bibinfo {year}
  {2001})}\BibitemShut {NoStop}%
\bibitem [{\citenamefont {Hutchinson}(1990)}]{StochEst}%
  \BibitemOpen
  \bibfield  {author} {\bibinfo {author} {\bibfnamefont {M.}~\bibnamefont
  {Hutchinson}},\ }\bibinfo {title} {{A stochastic estimator of the trace of
  the influence matrix for Laplacian smoothing splines}},\ \href {\doibase
  10.1080/03610919008812866} {\bibfield  {journal} {\bibinfo  {journal}
  {Commun. Stat. - Simul. Comput.}\ }\textbf {\bibinfo {volume} {19}},\
  \bibinfo {pages} {433} (\bibinfo {year} {1990})}\BibitemShut {NoStop}%
\bibitem [{\citenamefont {Gropp}\ \emph {et~al.}(1996)\citenamefont {Gropp},
  \citenamefont {Kaper}, \citenamefont {Leaf}, \citenamefont {Levine},
  \citenamefont {Palumbo},\ and\ \citenamefont {Vinokur}}]{TDGL1}%
  \BibitemOpen
  \bibfield  {author} {\bibinfo {author} {\bibfnamefont {W.~D.}\ \bibnamefont
  {Gropp}}, \bibinfo {author} {\bibfnamefont {H.~G.}\ \bibnamefont {Kaper}},
  \bibinfo {author} {\bibfnamefont {G.~K.}\ \bibnamefont {Leaf}}, \bibinfo
  {author} {\bibfnamefont {D.~M.}\ \bibnamefont {Levine}}, \bibinfo {author}
  {\bibfnamefont {M.}~\bibnamefont {Palumbo}}, \ and\ \bibinfo {author}
  {\bibfnamefont {V.~M.}\ \bibnamefont {Vinokur}},\ }\bibinfo {title}
  {{Numerical simulation of vortex dynamics in type-II superconductors}},\
  \href {\doibase 10.1006/jcph.1996.0022} {\bibfield  {journal} {\bibinfo
  {journal} {J. Comput. Phys.}\ }\textbf {\bibinfo {volume} {123}},\ \bibinfo
  {pages} {254} (\bibinfo {year} {1996})}\BibitemShut {NoStop}%
\bibitem [{\citenamefont {Hansen}\ \emph {et~al.}(2006)\citenamefont {Hansen},
  \citenamefont {Kilmer},\ and\ \citenamefont {Kjeldsen}}]{HansenFFT}%
  \BibitemOpen
  \bibfield  {author} {\bibinfo {author} {\bibfnamefont {P.~C.}\ \bibnamefont
  {Hansen}}, \bibinfo {author} {\bibfnamefont {M.~E.}\ \bibnamefont {Kilmer}},
  \ and\ \bibinfo {author} {\bibfnamefont {R.~H.}\ \bibnamefont {Kjeldsen}},\
  }\bibinfo {title} {{Exploiting residual information in the parameter choice
  for discrete ill-posed problems}},\ \href {\doibase
  10.1007/s10543-006-0042-7} {\bibfield  {journal} {\bibinfo  {journal} {BIT
  Numer. Math.}\ }\textbf {\bibinfo {volume} {46}},\ \bibinfo {pages} {41}
  (\bibinfo {year} {2006})}\BibitemShut {NoStop}%
\end{thebibliography}%

\newpage
\renewcommand{\baselinestretch}{1}

\end{document}